\documentclass[prx,aps,twocolumn,floatfix,showpacs,10pt]{revtex4-1}
\usepackage{amsmath}
\usepackage{graphicx}
\usepackage{epsfig}
\usepackage{graphics}
\usepackage{bm}
\usepackage{amssymb}
\usepackage{psfrag}
\usepackage{mathtools}
\usepackage[english]{babel}
\usepackage[T1]{fontenc}
\usepackage{color}
\usepackage{hyperref}
\usepackage{ulem}
\usepackage{braket}
\usepackage{umoline}

\begin{document}
\title{Toolbox for Abelian lattice gauge theories with synthetic matter}
\author{Omjyoti Dutta$^{1,2,4}$}
\email{omjyoti@gmail.com}
\author{Luca Tagliacozzo$^{3,4}$}
\author{Maciej Lewenstein$^{4,5}$}
\author{Jakub Zakrzewski$^{1,6}$}
\affiliation{ \mbox{$^1$ Instytut Fizyki imienia Mariana
Smoluchowskiego, Uniwersytet Jagiello\'nski, ulica \L{}ojasiewicza
11, PL-30-348 Krak\'ow, Poland.} 
\mbox{$^2$ Donostia International Physics Center DIPC, Paseo Manuel de Lardizabal 4, 20018 Donostia-San Sebasti\'{a}n, Spain.}
\mbox{$^3$ Department of Physics and SUPA, University of Strathclyde, Glasgow G4 0NG, United
Kingdom} 
\mbox{$^4$ ICFO - Institut de Ciencies Fotoniques, The
Barcelona Institute of Science and Technology, 08860 Castelldefels (Barcelona), Spain.} 
\mbox{$^5$ ICREA, Lluis Companys 23, 08010 Barcelona, Spain.} 
\mbox{$^6$ Mark Kac Complex
Systems Research Center, Uniwersytet Jagiello\'nski, Krak\'ow,
Poland.}}

\date{\today}

\begin{abstract}
{Fundamental forces of Nature are described by field theories,
also known as gauge theories, based on a local gauge invariance.
The simplest of them is quantum electrodynamics (QED), which is an
example of an Abelian gauge theory. Such theories describe the
dynamics of massless photons and their coupling to matter.
However, in two spatial dimension (2D) they are known to exhibit gapped phases at low
temperature. In the realm of quantum spin systems, it remains a
subject of considerable debate if their low energy physics can be
described by emergent gauge degrees of freedom. Here we present a
class of simple two-dimensional models that admit a low energy description in terms of an
Abelian gauge theory. We find rich phase diagrams for
these models comprising exotic deconfined phases and gapless phases  -
a rare example for 2D Abelian gauge theories. The counter-intuitive presence of gapless
phases in 2D  results from the emergence of additional symmetry in
the models. Moreover, we propose schemes to realize our model with 
current experiments using ultracold bosonic atoms in optical
lattices}
\end{abstract}

\maketitle
\section{Introduction}
The invariance of a system under the action of a local symmetry is one of the central ideas in modern
physics. It is the building block of gauge invariant theories that  provide the foundations of our current understanding of fundamental
forces of nature. The simplest of them, quantum electrodynamics,
describes the dynamics of gauge bosons (photons) and their
coupling to matter (electrons). 
Gauge theories can also emerge as a low energy description of low-dimensional quantum spin system \cite{fradkin1, wen1, lee}. Quantum spin liquids \cite{and},  exotic states of matter that do not break any symmetry down to zero temperature, provide the obvious example but the phenomenon is more general at least at a mean-field level.
Quantum fluctuations, however,  most of the time conspire against emerging gauge theories and strongly bind
the gauge bosons  together, producing a low
energy sector dominated by standard spin fluctuations \cite{kog1, horn, drell, banks}. For this
reason the validity of such  emergent gauge theory  descriptions
still remains a subject of  considerable debate \cite{lee, param,
fish, muramatsu, sorella, mot}.

In the present paper we take a fresh look on the emergence of gauge theories in the context of bosonic Hamiltonians in two dimensions.
We focus on Hamiltonians that can be realized in experiments with ultra-cold atoms, and describe in detail how to design the corresponding  experiments.
In particular we analyze a system made by two-species of bosons hopping on a two dimensional
square lattice.  Our main result shows that the emerging  gauge theory description of the system naturally
accounts for the appearance of an exotic gapless dipolar liquid phase. 

In particular we show how to design  the tunneling geometry of one of the
two bosonic species (referred to as the auxiliary particles), such that the other species is forced to  behave, at sufficiently low energies, as an effective gauge boson. We also describe the regime in which we can ensure that the gauge bosons remains mass-less even after including the quantum fluctuations.
The construction we propose is very flexible, and provides a complete toolbox for generating ``exotic'' low energy gauge theories.
The emerging gauge theories are always Abelian, but depending on the setup can have a discrete $\mathbb{Z}_N$ or continuous $U(1)$ symmetry.

Their Hamiltonians, however, are different from those used to describe high-energy ``standard'' gauge theories and, as a consequence, their phase diagrams are richer. Standard gauge theories with discrete gauge invariance (e.g. $\mathbb{Z}_N$) show both confined and deconfined phases \cite{horn,drell,kog1,banks}. Both are gapped,
and the extent of the latter in the parameter space vanishes as $N$ tends to infinity. As a result, the standard $U(1)$ gauge theory 
exhibits only the gapped confined phase, as was originally pointed
out by Polyakov \cite{Polyakov1,Polyakov2}. Our ``exotic'' gauge theories exhibit, in addition,  exotic gapped deconfined phases with dipolar and kink excitations and, most remarkably, a gapless dipolar liquid phase.

Taking a  complementary perspective we can identify the proposed bosonic system as a very flexible toolbox to generate, at low energies, exotic gauge theories in cold atom experiments. 
 Most of the seminal proposals in this direction have focused on implementing microscopic models of gauge theories, \cite{buch, edina, reznik, cirac0, luca,
zoller1, cirac1, lew1, reznik2, wiesse, zohar}. Instead, here we focus on  engineering emerging gauge theories rather than microscopic gauge theories. As a result we believe that our paper provides a plausible novel and original proposal to perform 2D quantum simulations of a gauge theories with ultra-cold atoms. 

\section{The model}
We consider a 2D lattice and  identify its sites  by Latin
letters ${\mathbf j} \equiv (j_x,j_y)$ and links by the  
pairs $({\mathbf j},{\hat \delta})$, where ${\mathbf j}$ denotes the site they originate from and ${\hat \delta}$ the direction they point to. It is understood that
the lattice direction is given by 
${\hat \delta}= \hat{x}\ {\mathrm or}\  \hat{y}$. With a
slight abuse of notation we also identify the nearest neighbor of
site ${\mathbf j}$ in the direction ${\hat \delta}$ as site ${\mathbf j}+{\hat \delta}$. On each
site of the lattice we have two species of bosons:
\emph{auxiliary} {\it  a}-bosons and {\it b}-bosons. The creation
and annihilation operators for the two species are respectively
$\hat{a}^{\dagger}_{\mathbf j}$, $\hat{a}_{\mathbf j}$ and $\hat{b}^{\dagger}_{\mathbf j}$,
$\hat{b}_{\mathbf j}$. The number operator for {\it b}-bosons is defined
as $\hat{n}_{\mathbf j}=\hat{b}^{\dagger}_{\mathbf j}\hat{b}_{\mathbf j}$. We assume that
{\it a}-bosons are hardcore, and thus we can have at most one {\it
a}-boson per site. We also  define the difference of the number
operators of {\it b}-bosons  on neighboring sites,
$\hat{n}_{({\mathbf j}, {\hat \delta})}\equiv \hat{n}_{{{\mathbf j}+{\hat \delta}}}-\hat{n}_{\mathbf j}$, and associate it to 
an operator living on the link $({\mathbf j},{\hat \delta})$ that connects the two sites.

\begin{figure*}
\includegraphics[width=\textwidth]{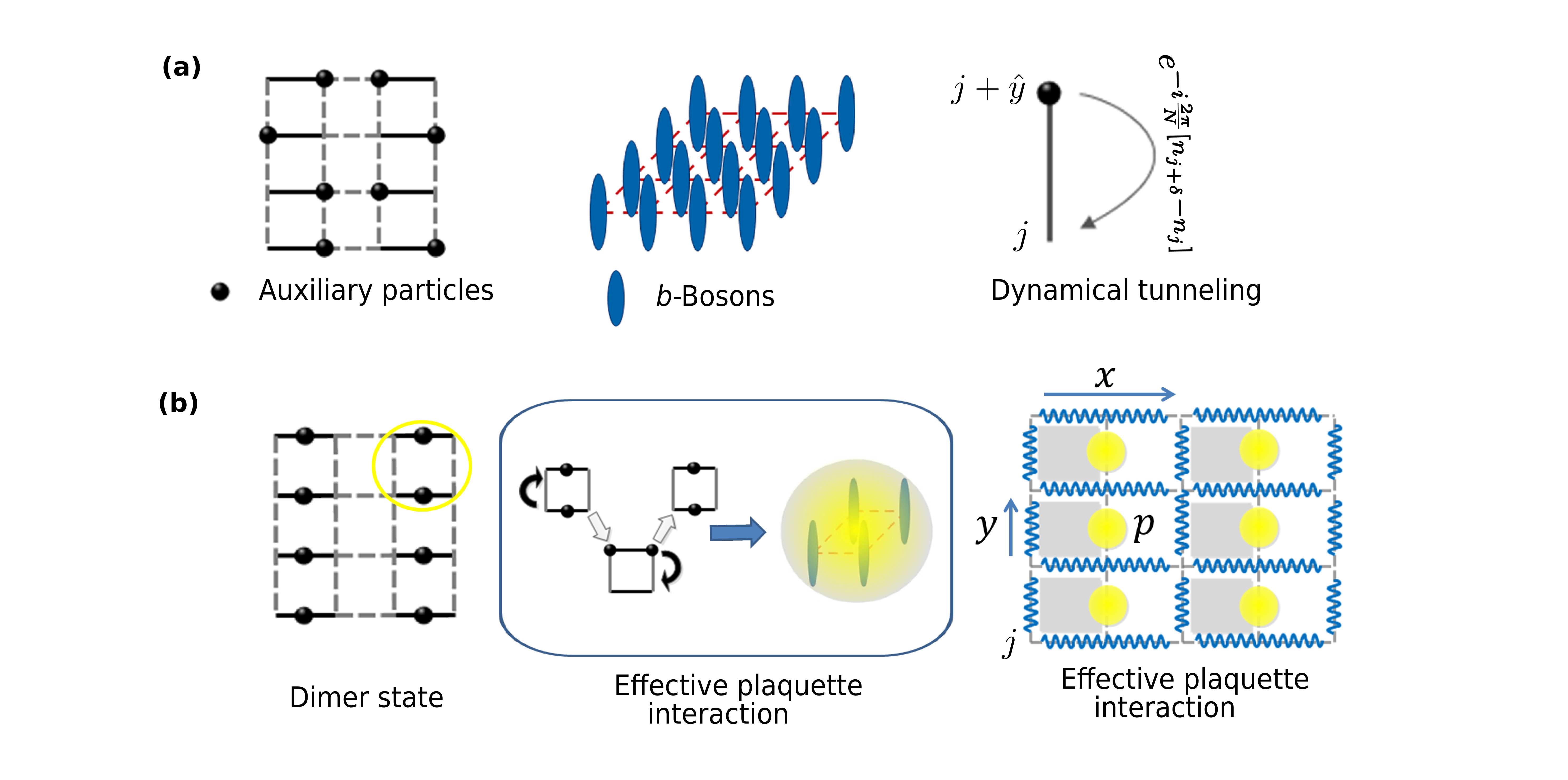}
\caption{Key ingredients - (a) \textbf{Left panel:} The auxiliary particles (black dots) are trapped in super-lattice geometry following the pattern in the figure. The tunneling rate is large (small) along dark (dashed) bonds. There is one particle for every dark bond. \textbf{Middle panel:} We need a large occupation of {\it b}-bosons on every site. This is achieved by trapping them along one-dimensional tubes (blue ovals) arranged in a square lattice geometry. \textbf{Right panel:} Upon appropriately shaking the set-up (see text for details),  the effective tunneling of the {\it a}-bosons, at low-frequencies,  is modulated  in phase by the presence  of the {\it b}-bosons. (b) \textbf{Left panel:} By further increasing the tunneling along dark bonds, the {\it a}-bosons delocalize on that bond. \textbf{Middle panel:} At second order in perturbation theory with respect to the weak tunneling, the virtual processes depicted create an effective plaquette interaction for the {\it b}-bosons (yellow sphere). \textbf{Right panel:} We can thus change variable to plaquette variables, that can be thought as belonging to a coarse-grained lattice (shown by the wiggly blue lines) where the electric fields and the vector potentials live.}
 \label{figure1}
\end{figure*}

Our aim is to investigate the low energy physics of the Hamiltonian
\begin{eqnarray}
\label{hameff}
H &=& -\sum_{{\mathbf j},{\hat \delta}} \left( J_a({\mathbf j},{\hat \delta})\hat{a}^{\dagger}_{{\mathbf j}} \exp\left[i\alpha_{{\hat \delta}}\hat{n}_{({\mathbf j},{\hat \delta})} \right]\hat{a}_{{\mathbf j}+{\hat \delta}} + h.c. \right )\nonumber
 \\
 &-&J_b\sum_{{\mathbf j},{\hat \delta}} \left(\hat{b}^{\dagger}_{{\mathbf j}} \hat{b}_{{\mathbf j}+{\hat \delta}} + h.c. \right )+  \frac{U}{2}\sum_{\mathbf j}(\hat{a}^{\dagger}_{{\mathbf j}})^2\hat{a}^2_{{\mathbf j}}  ,
\end{eqnarray}
where {\it h.c.} stands for Hermitian conjugate. This is a generalization of the Bose-Hubbard Hamiltonian, 
in which the tunneling amplitudes of {\it  a}-bosons are link dependent, $J_a({\mathbf j},{\hat \delta})$,  while those for {\it b}-bosons $J_b$ are uniform. More importantly, the phases of the tunneling amplitudes of {\it  a}-bosons are modulated by the occupation of  {\it b}-bosons on neighboring sites as illustrated in the right-hand panel of Fig.\ref{figure1}a.  The strength of the modulation, $\alpha_{{\hat \delta}}$, depends on the direction. It vanishes along the $x$ direction $\alpha_{x}=0$, while 
$\alpha_{y}=\alpha=2\pi/N$
 (with $N$ being a positive integer).
 The lattice configurations for both bosons are shown in Fig.\ref{figure1}a. 
The {\it b}-bosons are assumed to be non-interacting. The {\it a}-bosons interact strongly with $U$ being the dominant energy scale, so that in effect they are considered to be hard core bosons.    

In the present paper we show that the low energy sector of Hamiltonian \eqref{hameff}, is described by an exotic gauge theory for {\it b}-bosons filling $\bar{n} \gg 1 $. Note that similar models have been considered recently in  \cite{keilmann11,edmonds13,greschner14,arek, dutta15,greschner15,santos1,anna15,biedron16,bermudez15,david1}. We 
discuss the details of the physical implementations of the Hamiltonian (\ref{hameff}) to the later sections.
Let us here just mention that the above Hamiltonian can be engineered experimentally in an ultra-cold atoms set up. In that context the {\it a}-bosons tunneling inhomogeneity can be realized by trapping them in optical super-lattices. The large filling of {\it b}-bosons is obtained by allowing  the {\it b}-bosons to form extended tubes \cite{seng0} along a direction perpendicular to the lattice. Such species dependent traps have been demonstrated experimentally by using state-selective optical lattices \cite{bloch, hem, ess, seng1,seng2,seng}. The last ingredient is the possibility to tune both interaction among {\it a}-bosons and {\it b}-bosons, at the same time ensuring that {\it b}-bosons do not interact. This requires an accurate choice of the species representing the bosons.  For example one could choose  $^{39}$K-$^{133}$Cs mixture that has the 
necessary hyperfine structure accompanied by a rich landscape of Feshbach resonances \cite{patel}. One may use the inter-species resonance around $350$G \cite{patel} where the {\it a}-bosons interact strongly ($a_s\sim 2000$a$_0$) whereas {\it b}-bosons are essentially noninteracting ($a_s\sim 4$a$_0$). Still, since we assume large occupation of {\it b}-bosons per site, $\bar{n} \gg 1 $, one should tune the magnetic field to reduce  {\it b}-bosons interaction to an almost exact zero of the Feshbach resonance.

We present in the next Section \ref{gaugesec} our main result -- the  dynamical gauge theory and 'exotic' phases emerging from the Hamiltonian Eq.\eqref{hameff}. In Section \ref{toolbox}, we discuss the possible use of our setup as a toolbox for certain kind of gauge theories. In Section \ref{Implementation} we provide two different detailed schemes that use  periodic shaking mechanisms of the optical lattice and allow to generate the Hamiltonian, Eq.~\eqref{hameff}, as the effective Hamiltonian after time-averaging. 

\section{Emerging lattice gauge theory}
\label{gaugesec}

As an example of a   possible gauge theory we consider {\it a}-bosons hopping on a dimerized lattice as presented in the left-hand panel of Fig.~\ref{figure1}b. The considered tunneling amplitudes are  $J_a({\mathbf j},\hat{x})=J_{1x} >0 $ if $j_x$ is odd and $J_a({\mathbf j},\hat{x})=J_{2x} >0 $ for even $j_x$. The tunneling of auxiliary bosons along y-direction is given by $J_a({\mathbf j},\hat{y})=J_y >0$. We are interested in the limit where sites ${\mathbf j}$ and ${\mathbf j}+\hat{x}$ are dimerized for odd $j_x$, i.e $J_{1x} \gg J_{2x},J_y$. The {\it a}-bosons are in the insulating phase when tunnelings between the dimer links vanish, i.e $J_{2x}=J_y=0$. The insulating phase is represented by each dimerized link containing exactly one {\it a}-boson as in Fig.~\ref{figure1}b -- we assume here ``half-filling'' for the  {\it a}-bosons -- so that in the low energy sector  the $U$ term assures their hard-core nature. This gives our ``zero-order'' Hamiltonian:
\begin{equation}
\label{hamefforder0}
H_0= -J_{1x}\sum_{j_x,j_y} \left( \hat{a}^{\dagger}_{{\mathbf j}} \hat{a}_{{\mathbf j}+{\hat x}} + h.c. \right ), j_x \in {\rm odd}
\end{equation}

The ground state of this model, with energy denoted as $E_0$, is a Mott insulator of the {\it a}-bosons in the dimer states localized on the odd horizontal links, and an arbitrary state of the {\it b}-bosons. Being independent of the state of the {\it b}-species, it is thus highly degenerate. We denote the projector on the manifold of the degenerate ground states as $P$.  

 Now,  we take into account the {\it b}-bosons tunneling,  as well as $y$-tunneling ($J_y \ll J_{1x}$) within the perturbation theory. We will assume that the horizontal even bonds are strictly zero, $J_{2x}=0$. The perturbation consists of the diagonal
 and non-diagonal terms
\begin{equation}\label{hamefforder1}
H_1=H_D+H_{ND},
\end{equation}
where the diagonal part (i.e. the part,  which acts as a block matrix on the ground state manifold), and the non-diagonal part (i.e. the part that transforms the ground states outside the ground state manifold) are, respectively:  
\begin{eqnarray}\label{hamefforder1a}
H_D&-&J_b\sum_{{\mathbf j},{\hat \delta}} \left(\hat{b}^{\dagger}_{{\mathbf j}} \hat{b}_{{\mathbf j}+{\hat \delta}} + h.c. \right ),\nonumber\\
H_{ND}& =& -J_y\sum_{{\mathbf j}} \left(\hat{a}^{\dagger}_{{\mathbf j}} \exp\left[i\frac{2\pi}{N}\hat{n}_{({\mathbf j},{\hat y})} \right]\hat{a}_{{\mathbf j}+{\hat y}} + h.c. \right ).
\end{eqnarray}
The off-diagonal part mixes
  different dimer links, creating an effective potential for {\it b}-bosons due to phase modulation of the inter-dimer tunnelings.  
We could  also take into account the coupling between dimer rows due to small nonzero $J_{2x}$, but  at second order of the perturbation theory, this   leads to an uninteresting  constant term only.
Let us write the effective Hamiltonian to second order of the perturbation theory
  \begin{equation}\label{effectiveP}
  H_{eff}= H_D-PH_{ND}\frac{1}{H_0-E_0}H_{ND}P.
 \end{equation}
It is interesting to notice that at this level $H_D$ acts as a  \emph{kinetic} term for {\it b}-bosons, whereas  $PH_{ND}\frac{1}{H_0-E_0}H_{ND}P$ acts as a \emph{potential} term $H_{\rm pot}$.
 
Let us first discuss this potential term that is a genuinely  second order term in (\ref{effectiveP}). It depends on a linear combination of occupations around the shaded plaquettes $p$ (see Fig. 1b), containing the sites ${\mathbf j},{\mathbf j}+\hat{x},{\mathbf j}+\hat{x}+\hat{y},{\mathbf j}+\hat{y}$ for odd $j_x$. 
It is given by the sum of the shaded plaquettes , 
  \begin{equation}\label{eq:hampot}
H_{\rm pot}=-2K\sum_p \cos {\left[\hat{\mathcal{B}}_{p}\right]},
  \end{equation}
where the plaquette operators $\hat{\mathcal{B}}_p$ are introduced as 
  \begin{equation}\label{eq:plach}
\hat{\mathcal{B}}_{p} \equiv\frac{2\pi}{N}\left(\hat{n}_{\mathbf j} - \hat{n}_{{\mathbf j}+\hat{x}} + \hat{n}_{
{\mathbf j}+\hat{x}+ \hat{y}} - \hat{n}_{{\mathbf j}+\hat{y}}\right),
 \end{equation}
and plaquette strength $K = 2J^2_{b}/J_{1x} $. The $\hat{\mathcal{B}}_p$ operator has eigenvalues $\mathcal{B}_p \in (2\pi/N)\left[-(N-1)/2,..-1,0,1,....(N-1)/2\right]$. 

At this moment, $H_D$ in  the effective Hamiltonian is expressed in terms of the original creation and annihilation operators for the {\it b}-bosons, whereas the second order part is expressed by the plaquette operators for the shaded plaquettes. It is thus interesting to express $H_D$ in terms of the conjugated lattice gaige theory operators. 
To this aim we follow the  standard procedure \cite{drell}, and introduce the ladder operators, $\mathcal{L}_p$,  to construct a $\mathcal{Z}_N$ algebra.  $\mathcal{L}_p$ fulfills the following commutation  relations with the operator $\hat{\mathcal{B}}_p$ defined on the same plaquette $p$,
\begin{align} \label{eq:zn}
\left [\hat{\mathcal{L}}_p,e^{\mp i\hat{\mathcal{B}}_p } \right ] = \pm e^{ \mp i \hat{\mathcal{B}}_p },  \nonumber \\
\left [\hat{\mathcal{B}}_p, e^{\pm i \frac{2\pi}{N}\hat{\mathcal{L}}_p } \right ] = \pm\frac{2\pi}{N}e^{ \pm i \frac{2\pi}{N}\hat{\mathcal{L}}_p}.
\end{align}
while  commuting with  $\hat{\mathcal{B}}_{p'}$ defined on different plaquettes, $p'$. 

Next we can try to express the tunneling Hamiltonian of the {\it b}-bosons, $H_D\equiv K g^2 H_{\textrm{kin}}$,  in terms of  the ladder operators, where we have introduced an coupling constant $g^2$ (defined later in terms of tunneling amplitude) to make better contact with the standard notation used in the context of gauge theories.
Generic examples of such expressions  are:
\begin{eqnarray} \label{tunn}
&& \hat{b}^{\dagger}_{\mathbf j} \hat{b}_{{\mathbf j}+\hat{x}}\left |\mathcal{B}_p,\mathcal{B}_{p-\hat{y}} \right \rangle;\quad {j}_x \in odd  \nonumber\\
&=& \sqrt{(n_{\mathbf j}+1){n}_{{\mathbf j}+\hat{x}}} \left |\mathcal{B}_p + \frac{4\pi}{N}, \mathcal{B}_{p-\hat{y}} - \frac{4\pi}{N} \right \rangle \nonumber\\
&\approx & \bar{n} e^{- i \frac{4\pi}{N} \left[ \hat{\mathcal{L}}_p - \hat{\mathcal{L}}_{p-\hat{y}} \right]} \left | \mathcal{B}_p, \mathcal{B}_{p-\hat{y}} \right\rangle, \nonumber\\
&& \hat{b}^{\dagger}_{\mathbf j} \hat{b}_{{\mathbf j}+\hat{y}}\left |\mathcal{B}_p,\mathcal{B}_{p-\hat{y}},\mathcal{B}_{p+\hat{y}}  \right \rangle  \nonumber\\
&=& \sqrt{(n_{\mathbf j}+1){n}_{{\mathbf j}+\hat{y}}} \left |\mathcal{B}_p + \frac{4\pi}{N}, \mathcal{B}_{p-\hat{y}} - \frac{2\pi}{N}, \mathcal{B}_{p+\hat{y}} -\frac{2\pi}{N} \right \rangle \nonumber\\
&\approx & \bar{n} e^{i \frac{2\pi}{N} \left[ \hat{2\mathcal{L}}_p - \hat{\mathcal{L}}_{p-\hat{y}} -\hat{\mathcal{L}}_{p+\hat{y}} \right]} \left | \mathcal{B}_p, \mathcal{B}_{p-\hat{y}},\mathcal{B}_{p+\hat{y}} \right\rangle,\\
&& \hat{b}^{\dagger}_{\mathbf j} \hat{b}_{{\mathbf j}+\hat{x}}\left |\mathcal{B}_p,\mathcal{B}_{p-\hat{y}} ,\mathcal{B}_{p+\hat{x}},\mathcal{B}_{p+\hat x-\hat{y}}\right \rangle;\quad {j}_x \in even  \nonumber\\
&\approx & \bar{n} e^{ i \frac{2\pi}{N} \left[ \hat{\mathcal{L}}_p + \hat{\mathcal{L}}_{p+\hat{x}}- \hat{\mathcal{L}}_{p-\hat{y}}- \hat{\mathcal{L}}_{p+\hat x-\hat{y}} \right]} 
\left |\mathcal{B}_p,\mathcal{B}_{p-\hat{y}} ,\mathcal{B}_{p+\hat{x}},\mathcal{B}_{p+\hat x-\hat{y}}\right \rangle, \nonumber
\end{eqnarray}
where the final expressions are valid only for a large {\it b}-boson filling, $\bar{n}\gg 1$. In this limit $H_D$ may be expressed fully in terms of plaquette ladder operators. 
\begin{figure*}
    \includegraphics[width=\textwidth]{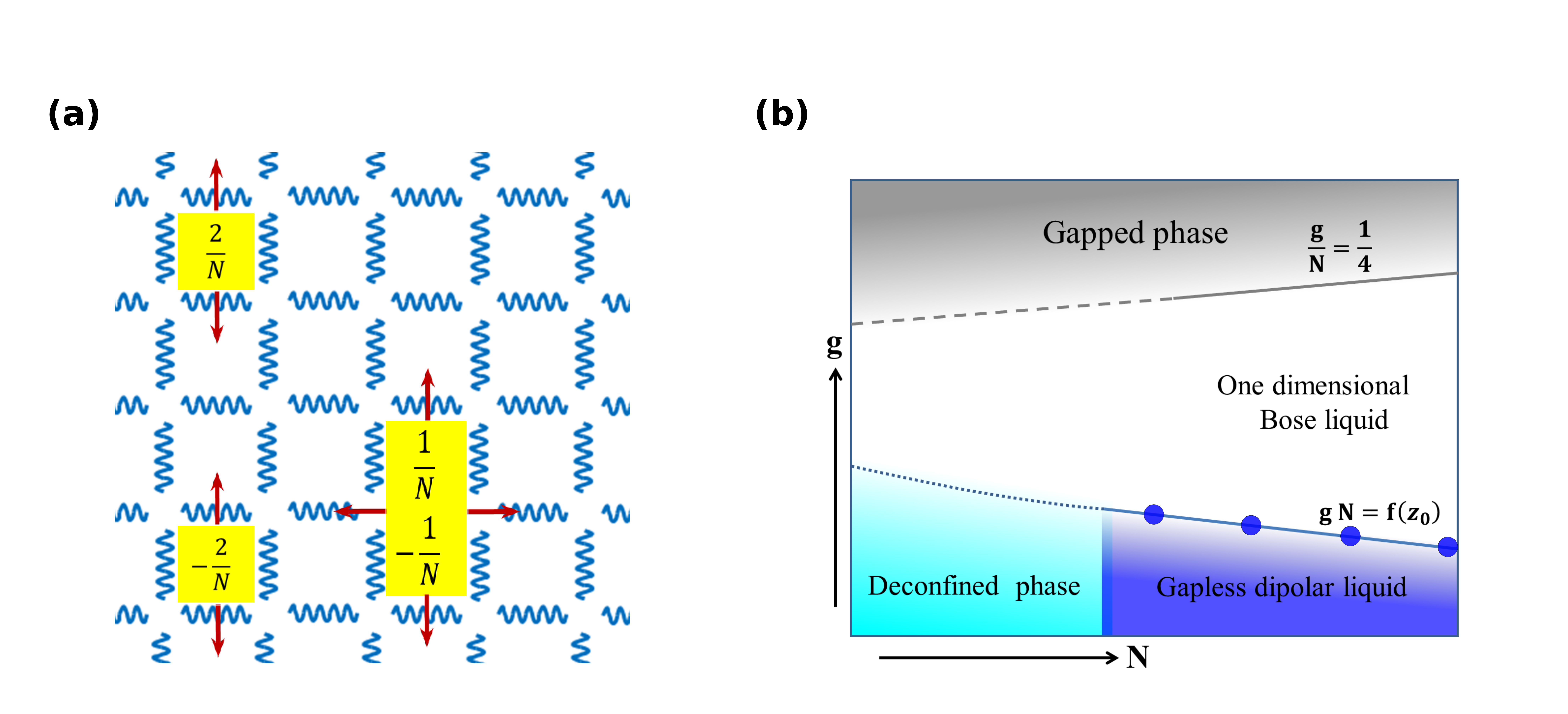}
    \caption{Low-energy excitations and phase diagram (a) At week coupling the low energy excitations are plaquette excitations, magnetic fluxes of strength $\pm 2/N$. Excitations can only be created in pairs inside a column and are free to move along that column, Alternatively, dipoles of magnetic fluxes involving the excitations of two adjacent plaquettes are free to move along both lattice directions. 
    (b) Qualitative phase diagram of the Hamiltonian $H_{\rm plaq}$ in the $g-N$ plane. The upper shaded region denotes a gapped phase in the strong coupling limit. In the weak coupling limit, for low $N$, the system is gapped but deconfined (region shaded in light blue). In the $U(1)$ limit, the system becomes gapless and a exotic dipole liquid phase emerge (region shaded in dark blue). In the intermediate region, the system is effectively in a one-dimensional gapless bose liquid phase.} 
        \label{figure2}
\end{figure*}

We also carry out the transformations: $\hat{\mathcal{L}}_{p}\rightarrow (-1)^{j_x}\hat{\mathcal{L}}_{p}, \hat{\mathcal{B}}_{p}\rightarrow (-1)^{j_x}\hat{\mathcal{B}}_{p}$. The resulting Hamiltonian reads in dimensionless units, i.e. after dividing by $K$:
\begin{equation}\label{plaqham0}
H_{\rm plaq}\equiv H_{eff}/K= (H_{ND} +H_D)/K = H_{\rm pot}/K+  g^2H_{\rm kin},
\end{equation} where
\begin{eqnarray}\label{plaqham1}
H_{\rm kin} &=& 2\sum_{p} \left[ \cos\left(\frac{4\pi}{N} \left[ \hat{\mathcal{L}}_{p}-\hat{\mathcal{L}}_{p -\hat{y}}\right] \right)  \right.  \nonumber\\
&+& \left. \cos\left(\frac{2\pi}{N} \left[\hat{\mathcal{L}}_{p}-\hat{\mathcal{L}}_{p+\hat{x}}- \hat{\mathcal{L}}_{p +\hat{y}}+\hat{\mathcal{L}}_{p +\hat{x} +\hat{y}}\right] \right) \right. 
\nonumber\\
 &+& \left.
 2 \cos\left(\frac{2\pi}{N}\left[2 \hat{\mathcal{L}}_{p}- \hat{\mathcal{L}}_{{p}+\hat{y}} - \hat{\mathcal{L}}_{p -\hat{y}}\right]\right) \right], 
\end{eqnarray}
and coupling strength $g^2=\bar{n}J_b/K$. 
 In \eqref{tunn}, the first, second and third terms lead  to first, third, and second tunneling expression in \eqref{plaqham1}, respectively. Note that the sums in the above Hamiltonian  run over the shaded  plaquettes. The interactions induced by $J_{1x}$
and all vertical bonds occur between the shaded plaquettes within each shade column independently. It is only the $J_b$ tunnelings that induce effective coupling between the shaded columns. This important asymmetry between horizontal and vertical direction is an key feature of our model, which in fact leads to its exotic properties.

The $\mathcal{Z}_N$ Hamiltonian \eqref{plaqham0} describes a spin model defined on the dual lattice whose sites are at the centers of plaquettes of the original lattice.  Therefore, by performing a duality transformation, we can go back to the original lattice and write the spin model as a gauge theory.
Since originally \eqref{plaqham1} involves only every second plaquette, the lattice, where the gauge theory is defined, is a coarse-grained version of the original lattice made by grouping  together $2$ plaquettes inside the $2\times 1$ squares as shown in Fig.~\ref{figure1}(b).
As a result, from now on, the plaquette index $p$ denotes the entire $2$ plaquettes group. We reverse the standard duality between gauge and spin systems \cite{fradkin2, horn, drell, fradkin1, savit,luca2} and define the electric-field operator  $\hat{\mathcal{E}}_{({\mathbf j},\hat{x})}=\hat{\mathcal{L}}_{p}-\hat{\mathcal{L}}_{p -\hat{y}}$ on the links of the coarse grained lattice.
Similarly $\hat{\mathcal{E}}_{({\mathbf j},\hat{y})}=-\hat{\mathcal{L}}_{{p}}+\hat{\mathcal{L}}_{p-\hat{x}}$. By construction, the electric-field operators obey the Gauss law: $\sum_{\delta} \hat{\mathcal{E}}_{({\mathbf j},\delta)}- \hat{\mathcal{E}}_{({\mathbf j}-\delta, \delta)}=0$.
The relation between $\hat{\mathcal{E}}$  and $\hat{\mathcal{L}}$ can be inverted in several ways, one possibility is $\hat{\mathcal{L}}_{p}= \sum^{n=0}_{n=-\infty}\hat{\mathcal{E}}_{({\mathbf j}-n \hat{x},\hat{y})}$.
Under the same duality transformation, the plaquette magnetic field is transformed to the standard curl of the Wilson line, $e^{i\hat{\mathcal{B}}_{p}} \equiv \hat{U}_{({\mathbf j},\hat{x})} \hat{U}_{({\mathbf j}+\hat{x},\hat{y})}\hat{U}^{\dagger}_{({\mathbf j}+\hat{y},\hat{x})} \hat{U}^{\dagger}_{({\mathbf j},\hat{y})}$ 
where the $\hat{U}$ operator has the same form as the  $\exp [i \hat{\mathcal{B}}]$ operator, but is defined on the links of the lattice. Since also $\hat{\mathcal{E}}$ has the same form as $\hat{\mathcal{L}}$,  $\hat{U}$ and $\hat{\mathcal{E}}$ fulfill the commutation relations \eqref{eq:zn} when acting on the same link and commute on different links. After the duality transformation, the Hamiltonian \eqref{plaqham0} becomes,
\begin{eqnarray}\label{eq:gaugeham}
{H_{\rm gauge}} &=& -2\sum_{p} \cos\hat{\mathcal{B}}_p - 2g^2 \sum_{\mathbf j} \left[ \cos\left ( \frac{4\pi}{N} \mathcal{E}_{({\mathbf j},\hat{x})}\right ) \right.\nonumber\\
&+& \left. 2\cos\left(\frac{2\pi}{N}[\mathcal{E}_{({\mathbf j},\hat{x})}-\mathcal{E}_{({\mathbf j}+\hat{y},\hat{x})}]\right)\right.\nonumber\\
&+&  \left.\cos\left(\frac{2\pi}{N}[\mathcal{E}_{({\mathbf j},\hat{y})}-\mathcal{E}_{({\mathbf j}+\hat{y},\hat{y})}]\right) \right],  
\end{eqnarray}
where the sums run now over the rectangular, coarse-grained plaquettes (see Fig. 1b). 

Expressing the low energy theory as a gauge theory gives us a better way to describe the phase diagram, and allows us to provide a precise prescription on how to measure the gauge field correlations in actual experiments. As an example, the $B$-field configurations can be obtained directly through counting the particle number at each sites. In the context of ultracold atoms, those can be easily available through single-site measurements \cite{ott, griener,bloch1} or by measuring the $\pi$-momentum particle density of a time-of flight image of a plaquette. Such correlations are important and perhaps the unique way to probe the phases of the model that as we will discuss in detail include a  Coulomb phase, where the gauge fields correlations decay as a  power law of their separation,  or the gapped spin-liquid phases where the gauge field correlations decay exponentially. Moreover, magnetic field correlations can clearly distinguish between the spin-liquid phases of the present work and standard superfluid or 
Mott-insulating phases, where $\langle \hat{\mathcal{B}}_{p+R} \hat{\mathcal{B}}_{p} \rangle = 0$ for $|R| \gg 1$.\\

\textbf{Weak coupling phases, $g \to 0$. }
When $g=0$, only the first term in Eq.~\eqref{eq:gaugeham} survives, and the ground state is given by a state with no magnetic charge, i.e, ${\mathcal{B}}_{p}=0,\  \forall p$. In the presence of a small non-zero coupling strength $g$, the  first excited states consist of frustrating two ${\mathcal{B}}_{p}$'s from the same column in order to fulfill charge conservation. Obviously, the phase  is gapped, since the excitations have a finite energy. As we shall see below, the nature of low energy excitations undergoes crossover from the magnetic charge to magnetic dipole excitations as $N$ grows.   

 We start by  constructing two families of  delocalized excitations. The first family contains two frustrated plaquettes in the same column having magnetic charge of unit $\pm 2/N$ as shown in Fig.\ref{figure2}(a).
An example of  the magnetic charge state is given by $\left | \pm  2/N,\mp 2/N \right \rangle_{p,p'}=   \exp\left(\pm i\frac{4\pi}{N}\hat{\mathcal{L}}_{p}\right ) \exp\left(\mp i \frac{4\pi}{N}\hat{\mathcal{L}}_{p'}\right )\left | 0\right \rangle$  where $p_x=p'_x$ so that the charges are from the same column and $|0\rangle$ is the state with no magnetic charge. Such magnetic charge states are created and de-localized by the action of the first row of the kinetic part of the Hamiltonian in Eq.\eqref{plaqham1}. 
It is thus natural to consider both $p,  p'$ in the same column $C$, and study the energy of the maximally de-localized state of a column  
$\ket{2}_C = \frac{1}{\sqrt{2 L(L-1)}} \sum_{p_x,p'_x\in C} \ket{\pm  2/N,\mp 2/N}_{p,p'}$, where $L \times L$ is the lattice size. Corresponding excitation energy is given by $2\Delta E_c$ with $ \Delta E_{c}=(1-\cos\left[4\pi/N\right])-2g^2$ for $L \rightarrow \infty$ .
The other family of states contains the  magnetic dipole state oriented along $y$: 
$$
\left| \uparrow_s \right\rangle = \exp\left(i \frac{2\pi}{N}\hat{\mathcal{E}}_{({\mathbf j},\hat{x})}\right ) \left | 0 \right \rangle.
$$
In particular we can again consider the zero momentum state $\left| \uparrow \right\rangle=\frac{1}{L(L-1)} \sum_{s} \left| \uparrow_s \right\rangle$ that has 
an excitation energy $\Delta E_{\uparrow}=2(1-\cos \left[2\pi/N\right]) - 6g^2$. Similarly, one could construct magnetic dipoles oriented along $-y$-direction  with the opposite sign in the exponent. As opposed to magnetic charges, the dipoles delocalize by tunneling along both lattice directions due to the last two terms in the Hamiltonian \eqref{eq:gaugeham} as pictorially sketched in Fig.~\ref{figure2}(b). 

\begin{figure}
	\includegraphics[scale=0.35]{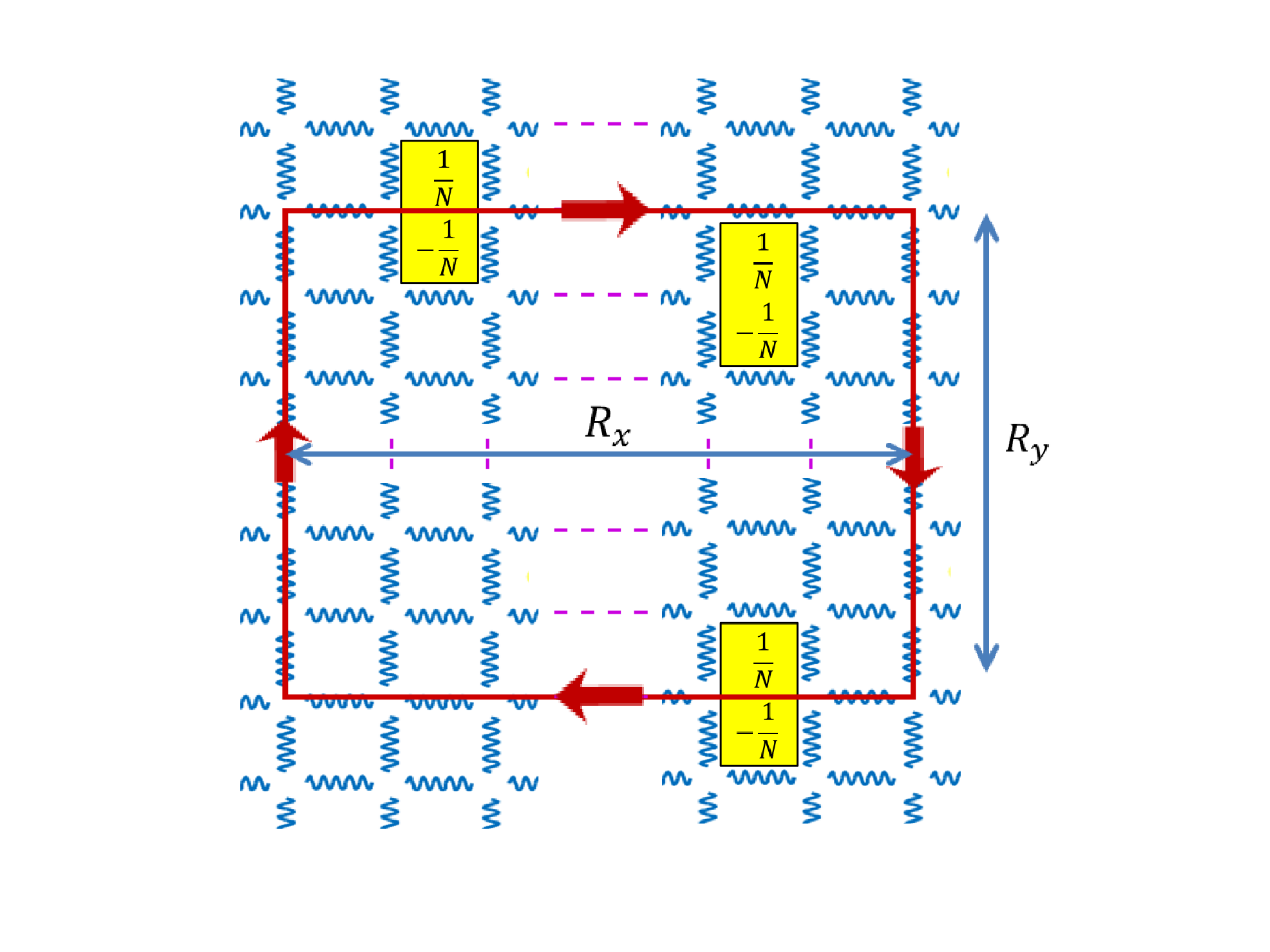}
	\caption{ A cartoon showing the Wilson loop in the coarse-grained lattice. The red line depicts the loop along which the vector potential is calculated. All the plaquettes inside the red line contribute to Wilson loop. Moreover, we have shown three possible positions for a single dipole. The top left and the bottom right dipole positions contributes to the Wilson loop a factor of $\exp(\pm i 2\pi/N)$. The dipole placed in the top right position is totally inside the loop area and has a unity contribution. } 
		\label{supfig1}
\end{figure}
Comparing $2\Delta E_c$ and  $\Delta E_{\uparrow}$, we see that for $N \le  3$ we have $2\Delta E_c\le \Delta E_{\uparrow}$ for a  weak coupling, and  the lower energy excitations are magnetic charge states; the ground state is here similar to the deconfined phase of the corresponding standard gauge theories. This can be seen by measuring the expectation value of the Wilson loop  \cite{wilson},
 \begin{equation}\label{eq:w_loop}
  W = \Pi_{({\mathbf j},\hat \delta)\in RC}\hat{U}_{({\mathbf j},\hat \delta)},
 \end{equation}
where $RC$ is a closed loop on the lattice, shown by the red tine in Fig.\ref{supfig1}. The deconfined phase is characterized by $\langle W \rangle \propto \exp[- P(RC)]$ where $P$ stands for the perimeter of the loop $RC$. This means that the expectation value of Wilson loop in the coarse-grained lattice decays to zero exponentially fast with the perimeter of the loop.

We now focus on the scenario where dipoles have lower energy than magnetic charges, which occurs for $N>3$.  Here we can write the ground state for small $g$ as $|g\rangle=|0\rangle - 3 g^2\left| \uparrow \right\rangle/(4\sin^2\left[\pi/N\right])$. As a consequence, the expectation value of the rectangular Wilson loop (see Fig.~\ref{figure2}(c)) ($RC$) of width $R_x$ and height $R_y$ is  $\langle W\rangle \sim \exp[- (9\cos\left[\frac{2\pi}{N}\right]/8) (g/\sin\left[\frac{\pi}{N}\right]) ^4 R_x] $.  We call this regime of the gapped deconfined phase {\it dipole-deconfined} since the first excited state consists of delocalized magnetic dipoles. A footprint of this phase is that the Wilson loop expectation value only decays exponentially fast in the horizontal width of the enclosed area.
Here we point out that as one goes from $N\leq 3$ to $N>3$, the ground state remains deconfined though the low-energy excitations change from being charge-like to dipole-like. \\

\textbf{Gapless Dipolar Liquid.}
In order to investigate the presence of a gapless phase, we focus on the nature of the system in the $U(1)$ limit of $N\rightarrow \infty, g \rightarrow 0$ with $gN=$constant. In this limit, we derive (details in Appendix \ref{A_p}) an equivalence of the partition function between Hamiltonian in Eq.~\eqref{eq:gaugeham} and dipolar Sine-Gordon model in Euclidean space-time (with renaming the co-ordinates as $x \equiv j_x, y\equiv j_y$, and $q \equiv (x,y,\tau)$): $Z_{\rm dipole}=\int D\varphi e^{-S_{\rm dipole}}$. The dipolar action $S_{\rm dipole}$ reads,
\begin{eqnarray}\label{sine-gordon}
S_{\rm dipole} &=& \int d^3q \Big[(\partial^2_{xy}\varphi_q)^2+(\partial^2_{yy}\varphi_q)^2 
+ (\partial_\tau \varphi_q)^2  \nonumber\\ 
&-& z_0 \cos\left(\frac{4\pi^{1/2}}{(gN)^{1/4}}\partial_y\varphi_q\right)\Big], 
\end{eqnarray}
where $z_0$ is the fugacity of dipole excitations, $\varphi$ denotes charge excitation field and the integral $\int D\varphi=\int\Pi_{q}d\varphi_q$. Additionally we introduce the symbolic differentials action on a function $f_q$ as, $\partial_{\epsilon} f_q= f_q - f_{q-\hat{\epsilon}}$ where $\epsilon=x,y,\tau$. Unlike the original sine-Gordon model \cite{frohlich}, the kinetic part of the field operators have quadratic and quartic components.
Moreover, the cosine potential contains a derivative of the fields along $y$-direction, originating from the presence of dipolar excitations
of the underlying model \cite{frohlich}. We solve Eq.\eqref{sine-gordon} variationally by using Gibbs-Bogoliubov-Feynman inequality. Our variational trial Gaussian action is given by expanding the cosine, $S_0=\int d^3q[(\partial^2_{xy}\varphi_q)^2+(\partial^2_{yy}\varphi_q)^2 
+ (\partial_\tau \varphi_q)^2 + m^2 (\partial_y\varphi_q)^2]$, where $m$ is a variational parameter. The Gibbs-Bogoliubov-Feynman inequality then states that $F_{\rm dipole}\leq F_{\rm var}=-T\langle\log Z_0\rangle_0 + T\langle S_{\rm dipole}-S_0\rangle_0$. When $m=0$, the propagator for the kinetic energy is quartic along the $x-y$ plane, whereas for non-zero $m$, the effective kinetic part of the action becomes quadratic along $\tau-y$ plane. The resulting free energy is expressed as, 
\begin{equation}
\mathcal{F}(m) \approx T\Big[\frac{1}{8\pi}|m|-2z_0\exp\left(-\frac{1}{4(gN)^{1/2}|m|}\right)\Big],
\end{equation}
where we have introduced a short-distance cutoff along $y$ direction $\sim 1$ corresponding to the lattice spacing and only considered the $m$ dependent terms. We find the optimum free energy by minimizing with respect to $m$. For each fugacity $z_0$, there exists a critical strength $(gN)_c$ such that for $(gN)<(gN)_c$, the optimum free energy is obtained for $m=0$, whereas for $(gN)>(gN)_c$, optimal $m$ is non-zero.

We term the phase with $m=0$ as a {\it dipolar liquid phase} with exotic correlation functions. For example, the charge-charge correlation function $C(q)=\langle \exp[i\varphi_q-i\varphi_0]\rangle=\delta_{x,0} C(y,\tau)$ is given by $ C(y,\tau) \sim \left[\frac{1}{y}\right]^{\eta(gN)\log y }$ when $ y \gg \tau \gg \sqrt{\tau} \gg 1 $, where $\eta(gN)$ is a function of coupling strength $gN$. The correlation function has an intermediate character between a power-law decay and an exponential decay. Moreover, the dipolar correlation function $D(q)=\langle \exp[i\partial_y\varphi_q-i\partial_y\varphi_0]\rangle $, has a three dimensional character. As an example, we find that, $D(0,y,\tau) \sim y^{\eta_1(gN)/y^2 }$ when $ y \gg \tau \gg 1 $ and $D(x,0,0) \sim x^{\eta_2(gN)/x }, x \gg 1 $, where $\eta_{1,2}(gN)$ are functions of coupling strength. More examples of such correlation functions are shown in Appendix \ref{A_c}.

\textbf{One dimensional Bose Liquid.}
In the one dimensional limit with $m\neq 0$, the charge-charge correlation functions are similar to a $1+1$ dimensional gapless {\it xy} model, $ C(y,\tau) \sim \left[\frac{1}{\tau^2+y^2}\right]^{\eta_4}$, and the corresponding dipolar correlations also show one dimensional character. To measure such correlations experimentally, we connect the charge correlation to the $\mathcal{L}$ operators $C(x,y,0)=\langle \exp \left[i\hat{ \mathcal{L}}_{p} - i\hat{ \mathcal{L}}_{0}\right] \rangle $. The scaling of these correlation functions, in principle, can be obtained from the experimental analysis of the visibility of interference fringes, similar to the one already performed in \cite{essbloch}.\\

\textbf{Strong coupling phases, $g \to \infty$ : }
We start with the $g \to \infty$ limit in Eq. \eqref{eq:gaugeham}.
In this limit the low energy sector is highly degenerate. Indeed any choice of constant $\mathcal{L}$ in the $\hat{y}$ direction, $\mathcal{L}_{p}=\mathcal{L}_{p+\hat{y}}$ minimizes \eqref{eq:gaugeham} at $g \to \infty$. This means that in a $L \times L$ lattice we have $N^L$ degenerate ground states. We can label these states by the value of the constant  $\hat{\mathcal{L}}$ operator of the plaquettes in a given column $p_x$ $\ket{\mathcal{L}}_{p_x}$ and see that the ground state becomes
\begin{equation}
 \set{\ket{\Omega}} =\set{\prod_{p_x} \ket{\mathcal{L}}_{p_x}} \quad \forall \ket{\mathcal{L}}_{p_x}.
\end{equation}
We can construct a column Fourier basis  $\ket{\mathcal{B}}_{p_x} = 1/\sqrt{3}\sum_{\mathcal{L}} \exp (i \mathcal{B} \mathcal{L}) \ket{\mathcal{L}}_{p_x}$.
In particular we can see that the ladder operator for the $\ket{\mathcal{L}}_{p_x}$ is given by $  \exp (i \hat{\mathcal{B}})_{p_x} =\prod_{p_y}  \exp (i \hat{\mathcal{B}}_{(p_x, p_y)})$. Under periodic boundary conditions, our magnetic field operator satisfies $\sum_C \mathcal{B}_p =0$. As a result, the ground state becomes non-degenerate and is given by, $\ket{\Omega} = \prod_{p_x}\ket{\mathcal{B}=0}_{p_x}$. The first excited state is again made of a manifold of states where one of the column state is changed from $\ket{\mathcal{L}}_{p_x}$ to $1/{\sqrt{L}}\sum_y \ket{\mathcal{L}^1 \mathcal{L}^2}_{p_x,y}$, where $\ket{\mathcal{L}^1 \mathcal{L}^2}_{p_x,y} = \prod_{p_y< y } \ket{\mathcal{L}^1}_{(p_x,p_y)} \otimes \prod_{p_y\ge y } \ket{\mathcal{L}^2}_{(p_x,p_y)}$ and $\mathcal{L}^1, \mathcal{L}^2$ are two different eigenvalues of the $\hat{\mathcal{L}}$ operator.
These states have a gap of order $1$ over the ground state manifold. Interestingly these excitations are localized in the $x$ direction, since once more their hopping only arises at order $L$ of the perturbation theory. Nevertheless there is no energy cost related to separating two of them in the $x$ direction so that they are energetically deconfined. The same holds in the $y$-direction, so that these domain walls are deconfined in both direction. Next one can show that, in the periodic QED limit \cite{drell,horn}, the system is equivalent to a $1+1$ dimensional {\it xy}-model, which has a gapped phase for $ g/N > 1/4$ (effective high temperature phase of the classical Coulomb gas). We present in Fig.~\ref{figure2}(b) a summary of the qualitative phase diagram.

\section{A toolbox for generating low energies gauge theories}
\label{toolbox}
In the previous section we have provided a specific atomic set-up whose low energy is described by a class of emerging gauge theories, all Abelian, from $\mathbb{Z}_2$ to $U(1)$ displaying exotic deconfined phases. This already provides a great deal of flexibility, since typically, changing the gauge group requires important changes in the implementation.  Here we want to explain that the setup we propose can be easily adapted to generate an even larger set of theories. For example, here we have focused on the specific case in which the auxiliary boson are in a Mott insulating phase and are integrated out from the low energy dynamics, but we could consider a different regime in which the auxiliary particle behave as fully dynamical charged matter fields. Furthermore we have chosen the dimerized configuration of auxiliary particles sketched in Fig. \ref{figure1}, but this can be generalized to, for example, configurations in which there is only one auxiliary particle every plaquette, giving rise to a 
different low energy theory. 
The bosons could also be trapped in different lattice geometries (for example triangular or honeycomb lattices rather than the square lattice considered here), once more giving rise to different low energy theories, appearing at different orders in perturbation theory. 
Finally, the auxiliary bosons can be substituted by auxiliary fermions, so to introduce a new energy scale, the Fermi energy, that could drastically modify the low-energy physics.  These are just few of the possible extensions we are currently characterizing. 
As a result, the set-up we are considering here constitutes a very flexible toolbox to generate gauge theories at low energy.

\section{Cold atoms implementation}
\label{Implementation}

Having described already the main results of the theory emerging from \eqref{hameff} let us discuss the details for the derivation of the effective Hamiltonian. Recall that we consider two species of particles, auxiliary, {\it a}-bosons and {\it b}-bosons. 
The assumed optical lattices potentials for these particles are
\begin{eqnarray} \label{bosdepth}
V^{a}_{\rm lat}&=&V^{a}_x \left [(1-S)\sin^2(\pi x/\lambda) + S \cos^2(2\pi x/\lambda)\right ] \nonumber\\
&+& V^{a}_y\cos^2(2\pi y/\lambda) + \frac{1}{2}m_a\Omega^2_{a}z^2 \\
V^{b}_{\rm lat}&=&V^{b}\left[\cos^2(\frac{2\pi x}{\lambda})+ \cos^2(\frac{2\pi y}{\lambda})\right] + \frac{1}{2}m_b\Omega^2_{b}z^2, 
\end{eqnarray}
 where $S$ parameter controls the relative heights of the super-lattice along $x$-direction for {\it a}-bosons. The optical lattice depths are denoted by $V^{\sigma}_{x,y}$. We assume a tight trap (with frequency $\Omega_{a}\gg \Omega_{b}$ ) for the {\it a}-particles. Here $\lambda$ is a typical optical laser wavelength. The masses of the {\it a}- and {\it b}-bosons are $m_a$ and $m_b$ respectively. The {\it b}-atoms are trapped in a square lattice with the lattice constant $\lambda/2$. In the third orthogonal direction {\it b}-bosons feel an elongated trap (compare the experimental setup in \cite{seng0}).  Such a scheme allows  to reach the regime of high boson fillings, namely: $\bar{n} \gg 1$. The superlattice potential enables dimerized tunnelings of auxiliary particles as discussed above (similar scheme has been implemented in \cite{Atala13}).

Let us now discuss  two possible implementations of shaking procedures that lead to the effective Hamiltonian Eq.~\eqref{hameff}.

\subsection{Scheme A: Shaking tunnelings and interactions}

We consider a standard tight-binding model in the lattice potential given  above with the Hamiltonian:
\begin{eqnarray}
\label{ham}
{H_{ab}} &=& -\sum_{{\mathbf j},{\hat \delta}}\left(J_a({\mathbf j},{\delta}) \hat{a}^{\dagger}_{\mathbf j} \hat{a}_{{\mathbf j}+{\delta}} + h.c. \right)+
\frac{U}{2}\sum_{\mathbf j}\hat{n}_{a{\mathbf j}}(\hat{n}_{a{\mathbf j}}-1) \nonumber \\
&-&J_b\sum_{\mathbf j,{\hat \delta}} \left(\hat{b}^{\dagger}_{\mathbf j} \hat{b}_{{\mathbf j}+{\hat \delta}} + h.c. \right)
+ U_{ab} \sum_{{\mathbf j}} \hat{n}_{a{\mathbf j}}\hat{n}_{b{\mathbf j}},
\end{eqnarray}
where we have added the inter-species interaction with strength $U_{ab}$ and the {\it b}-boson density is denoted by $\hat n_{bj}$ to better differentiate from the {\it a}-boson density $\hat n_{aj}$. Comparing \eqref{ham} with \eqref{hameff} we see that the phase modulation of the {\it a}-species tunneling amplitude is missing in \eqref{ham}. To generate the effective Hamiltonian,~\eqref{hameff} we add to $H_0$ periodically modulated (with frequency $\omega$) terms of the form

\begin{eqnarray}\label{shak_a}
H_{\rm sh} &=& - \cos\omega t \sum_{{\mathbf j},{\hat \delta}} \left(T_{{\hat \delta}}\hat{a}^{\dagger}_{\mathbf j} \hat{a}_{\mathbf j+{\hat \delta}} + h.c.\right)\nonumber\\
&+& U_{\rm sh}\sin\omega t \sum_{j} \hat{n}_{a\mathbf j} \hat{n}_{b{\mathbf j}},
\end{eqnarray}
where $T_{{\delta}}$ is a shaken component of the tunnelings and is non-zero only along the direction $T_{\hat{x}}=0, T_{\hat{y}}\neq 0 $. $U_{\rm sh}$ denotes the strength of the inter-species interaction modulation.

The harmonic shaking of the tunnelings may be realized by appropriately periodically modulating the depth of the
optical lattices in the $y$ direction:
\begin{equation}\label{shakamp_a}
V_{\rm a}(t)= V_{\rm sh} \cos\omega t \cos^2(2\pi y/\lambda),
\end{equation}
where $\omega$ is the shaking frequency and $V_{\rm sh}$ is  the amplitude modulation strength. The modulation of lattice depth not only induces the time dependence in tunneling but also induces periodic modulation in single-particle onsite energies. However, since the tunneling rates depend exponentially on the lattice depth the main effect of the lattice modulation is on tunneling rates. The shaking frequency, while large, should not be resonant with the energy difference between the $s$ and $p$ bands  \cite{lacki13}.

Observe that not only the tunnelings but also interactions between species are assumed to be modulated. The latter
can be performed with the help of magnetic Feshbach resonance (see below for discussion of possible choice of atomic species).

The Hamiltonian thus becomes $H(t)=H_{ab}+H_{\rm sh}$. To perform time-averaging over fast oscillations we
 apply the unitary transformation:  $\hat{U}=\exp [- { i} U_{\rm sh}  \sum_{{\mathbf j}} \hat{n}_{a{\mathbf j}} \hat{n}_{b{\mathbf j}} \int^t_0 \sin\omega t' dt']$ which transfers the time-dependence of the total Hamiltonian $H(t)$ into the tunneling amplitudes yielding $H_1=\hat{U}^\dagger H \hat{U} - {i}\hat{U}^\dagger [d_t \hat{U}]$. Using  Jacobi-Anger identity the Hamiltonian may be expressed as
\begin{eqnarray}
H_1 &=& H_{\rm a,av} + H_{\rm b,av}  + H_{\rm t} + H^{\dag}_{\rm t}, \\
H_{\rm a,av} &=& -\sum_{\mathbf j,{\hat \delta}} \hat{a}^{\dagger}_{\mathbf j} \mathcal{F}_{{\mathbf j\hat \delta}} \hat{U}_{{\mathbf j\hat \delta}} \hat{a}_{{\mathbf j}+{\hat \delta}} - h.c+\frac{U}{2}\sum_{\mathbf j}\hat{n}_{a{\mathbf j}}(\hat{n}_{a{\mathbf j}}-1)
\nonumber\\
H_{\rm b,av} &=& -J_{\rm b,av}\sum_{\mathbf j,{\hat \delta}} \hat{b}^{\dagger}_{\mathbf j} \hat{b}_{\mathbf j+{\hat \delta}} - h.c.\nonumber\\
H_{t}=&-& \sum_{\mathbf j,{\hat \delta}} J_{\rm a}({\mathbf j,\hat \delta})\sum_{M\neq 0} i^M \mathcal{J}_M(U_{\rm sh}\hat{n}_{b{\mathbf j{\hat\delta}}}/\omega)e^{iM\omega t}\hat{a}^{\dagger}_{\mathbf j} \hat{a}_{\mathbf j+{\hat \delta}}  \nonumber\\
&+& \cos \omega t\sum_{\mathbf j,{\hat \delta}}\sum_{M \neq \pm 1} T_{{\delta}} i^M\mathcal{J}_M(U_{\rm sh}\hat{n}_{b\mathbf j{\hat\delta}}/\omega)e^{iM\omega t}\hat{a}^{\dagger}_{\mathbf j} \hat{a}_{\mathbf j+{\hat \delta}}
\nonumber\\
&-&J_{\rm b}\sum_{\mathbf j,{\hat \delta}}\sum_{M\neq 0} i^M \mathcal{J}_M(U_{\rm sh}\hat{n}_{a{\mathbf j{\hat \delta}}}/\omega)e^{iM\omega t}\hat{b}^{\dagger}_{\mathbf j} \hat{b}_{\mathbf j+{\hat \delta}},\nonumber
\end{eqnarray}
with the tunneling amplitudes taking quite complicated form
\begin{eqnarray} \label{timeavg}
\mathcal{F}_{{\mathbf j\hat\delta}} &=& \sqrt{J^2_a({\mathbf j,\hat\delta})\mathcal{J}^2_0(U_{\rm sh}\hat{n}_{b{\mathbf j\hat\delta}}/\omega)+T^2_{{\hat\delta}}\mathcal{J}^2_1(U_{\rm sh}\hat{n}_{b{\mathbf j\hat\delta}}/\omega)} \nonumber\\
\hat{U}_{{\mathbf j\hat\delta}} &=& \exp\left[-i\tan^{-1}\left(\frac{T_{{\hat\delta}}\mathcal{J}_1(U_{\rm sh}\hat{n}_{b{\mathbf j\hat\delta}}/\omega)}{J_a({\mathbf j,\hat\delta})\mathcal{J}_0(U_{\rm sh}\hat{n}_{b{\mathbf j\hat\delta}}/\omega)}\right)\right], \\
{J}_{\rm b, av}&=&J_b\mathcal{J}_0(U_{\rm sh}\hat{n}_{a{\mathbf j\hat\delta}}/\omega).\nonumber
\end{eqnarray}
with, let us recall, $\hat{n}_{b{\mathbf j\hat\delta}}=\hat{n}_{b{\mathbf j+\hat\delta}} -\hat{n}_{b{\mathbf j}}$ being the difference of population operators for tubes separated by ${\hat\delta}$. While for each tube $\bar{n}>>1$, $\langle \hat{n}_{b{\mathbf j\hat\delta}} \rangle \ll \bar{n}$.
The tunneling of auxiliary particles is thus modulated dynamically both in amplitude \cite{dutta, santos1} and in phase by the presence of {\it b}-bosons.

The expressions \eqref{timeavg} simplify considerably assuming that the shaken tunneling component value
is chosen such that 
\begin{equation}\label{constun}
T_{\hat{y}}\approx \sqrt{2} J_y. 
\end{equation}
Then for
$U_{\rm sh} \bar{n}/\omega \lesssim 1$ one obtains the following approximate expressions:
\begin{equation}\label{approxphase}
\mathcal{F}_{{\mathbf j\hat \delta}} \approx J_a({\mathbf j,\hat\delta});\  \hat{U}_{{\mathbf j\hat\delta}} \approx \exp\left[-i\alpha_{{\hat\delta}} \hat{n}_{b{\mathbf j\hat\delta}}\right];\  {J}_{\rm b, av} \approx J_b.
\end{equation} 
Since the tunneling shaking is assumed along $y$ direction only we have
 $\alpha_{{\delta}_x}=0$, $\alpha_{{\delta}_y}=  U_{\rm sh}/\sqrt{2}\omega$.
 This approximation even for $U_{\rm sh} \bar{n}/\omega=1$ yields the error  less than ten percent. Within this approximation the {\it b}-boson tunneling remains unchanged. Furthermore, we assume  that $J_a, T_{\delta} \ll \omega$, so  $H_t$ contains fast oscillating terms only and may be dropped altogether (as it averages to zero over the period of the perturbation). Then we arrive at Eq.~\eqref{hameff}.

\subsection{Scheme B: Quasi-resonant lattice shaking}

The previous proposal allows us to make a weak modulation of tunneling phases.
To reach the regime of strong phase modulated tunnelings, we introduce another shaking scheme. We assume a standard lattice shaking potential \cite{Eckardt05,seng0} represented by the Hamiltonian,
\begin{eqnarray}
\label{shak_b}
H_{\rm sh}(t) &=&  K_b\cos\omega t \sum_{\mathbf j} (j_x+j_y) \hat{n}_{b\mathbf j}  \nonumber\\
&+& K \sum_{\mathbf j} \left[j_x\cos\omega t +
j_y\cos(\omega t + \phi) \right ]\hat{n}_{a\mathbf j} ,
\end{eqnarray}
where observe the additional phase difference in shaking for {\it a}-bosons. This is easily accomplished since we assume different lattices for both species anyway. As before we assume $\omega$ to be much larger than $J_a$ but
additionally we assume this frequency to be resonantly adjusted to the interspecies interaction strength $U_{ab}$
(or the latter to be modified by Feshbach resonance) with the condition:
\begin{equation}\label{resoU}
U_{ab}\approx  \mathcal{N}\omega
\end{equation}  with
$\mathcal{N}$ being an integer.

To average over the fast oscillations one has now to take this resonant condition into account. Define
$$H_r= \mathcal{N}\omega \sum_{{\mathbf j}} \hat{n}_{a{\mathbf j}} \hat{n}_{b{\mathbf j}}.$$ We transform the Hamiltonian to the rotating frame with the help of the unitary transformation taking the form
$$\hat U_2=\exp\left (-i\int_0^t H_{\rm sh}(\tau)d\tau-iH_rt\right).$$
where the first term takes care of the lattice shaking with frequency $\omega$ and strength $K$.

Again, in the limit of fast $\omega$ compared to other frequency (e.g. tunneling) scales, we carry out the time-averaging procedure, as for the first scheme, resulting in the effective Hamiltonian:
\begin{widetext}
\begin{eqnarray}\label{timeavgb}
H_1 &=& H_{\rm a} + H_{\rm b}+(U_{ab}-\mathcal{N}\omega ) \sum_{j} \hat{n}_{a\mathbf j}\hat{n}_{b\mathbf j},\qquad\quad H_{\rm b} = -J_{\rm b}\sum_{\mathbf j,{\hat\delta}} \mathcal{J}_{\mathcal{N}\hat{n}_{a{\mathbf j\hat\delta}}}( K_b/\omega) \hat{b}^{\dagger}_{\mathbf j} \hat{b}_{\mathbf j+{\hat\delta}} - h.c.,\nonumber\\
H_{\rm a} &=& -\sum_{\mathbf j,{\hat\delta}}\left[ J_a(\mathbf j,{\hat\delta}) e^{-i\beta_{{\hat\delta}}}\mathcal{J}_{\mathcal{N}\hat{n}_{b{\mathbf j\hat \delta}}}( K/\omega) \exp\left(i\alpha_{{\hat\delta}}\hat{n}_{b{\mathbf j\hat\delta}} \right) \hat{a}^{\dagger}_{\mathbf j}  \hat{a}_{\mathbf j+{\hat\delta}} +h.c.\right]
+\frac{U}{2}\sum_{\mathbf j}\hat{n}_{a{\mathbf j}}(\hat{n}_{a{\mathbf j}}-1)
\end{eqnarray}
\end{widetext}
where $\alpha_{{\delta_x}}=0, \alpha_{{\delta_y}}=\alpha=\mathcal{N} \phi$ and  $\beta_{{\delta_x}}=0,\  \beta_{{\delta_y}}=K\sin\phi/\omega$. The averaging procedure is valid provided
$\omega$ is much larger than the tunnelings and $U_{ab}-\mathcal{N}\omega\ll \omega$.

In the limit of strong shaking strength, $K$, and taking $\mathcal{N}=2$, we may approximate  Bessel functions as
\begin{equation}\label{aympbess}
\mathcal{J}_{2\hat{n}_{b{\mathbf j\hat\delta}}}(\frac{K}{\omega}) \approx \sqrt{\frac{2\omega}{\pi K}} \cos\left(\frac{K}{\omega}-\frac{\pi}{4}\right)\exp[-i\pi \hat{n}_{b{\mathbf j\hat\delta}} ],
\end{equation}
provided $K/\omega \gg 2{n}_{b{\mathbf j\hat\delta}}$. As a result we obtain the Hamiltonian of the form given by Eq.~\eqref{hameff} with modified, small $U_{ab}$, as well as renormalized $J_a(\mathbf j,{\hat\delta})$. For the tunneling of the {\it b}-bosons, we adjust the shaking parameter such that $ \mathcal{J}_0( K_b/\omega)=\mathcal{J}_2( K_b/\omega)$.

The effect of asymptotic approximation of the Bessel function is that the number fluctuations become constrained by the shaking strength. 
Let us also note that such a strong shaking  can induce heating due to the coupling to higher bands \cite{lacki13,andre} and the corresponding losses.
Those effects as well as possible choices of particular parameter values are discussed in Appendix \ref{A_r}.

\subsection{Necessary properties of atomic species}

The simulation of the gauge Hamiltonian requires noninteracting {\it b}-bosons with large mean occupation per site $n_b$, assured through appropriate lattice arrangement (with {\it b}-bosons confined to tubes perpendicular to the $Oxy$ plane. Thus one has  to choose almost non-interacting atomic species e.g. using the  zero crossing scattering length around Feshbach resonance. One experimentally available possibility is $^{39}$K in the hyperfine states $|1,1\rangle$ and $^{133}$Cs in the hyperfine state $|3,3 \rangle$ \cite{patel}. The role of auxiliary particles is taken by Cesium atoms which are hardcore bosons due to strong interactions. The magnetic field is tuned to a zero crossing of the Potassium atoms which is around $350$ Gauss \cite{Roati, lysebo, modugno}. This range of magnetic field is also suitable due to the presence of Feshbach resonance in K-Cs interaction at $340$ Gauss. Thus a time-periodic inter-species interaction around $350$ Gauss can be used to generate oscillating inter-species 
interaction (as required for the shaking scheme \textbf{A}). Moreover, one can of course additionally control the scattering length using optical Feshbach resonances \cite{grimm}, especially for generating the time-periodic force. Use of optical Feshbach resonance can potentially allow for a utilization  of a broader range of available ultracold atomic species.

Shaking scheme \textbf{B}, on the other hand is more versatile with respect to atomic species as no time modulation of interaction is necessary. The necessary condition is that the interspecies interactions, $U_{ab}$ are strong (again possible for the exemplary species discussed above close to the Feshbach resonance). Then adjusting the shaking frequency one can easily realize the resonant condition $U_{ab}=2\omega$ and with sufficiently strong shaking reach the regime described by the desired Hamiltonian, Eq.~\eqref{hameff}.

\section{Conclusions}
In conclusion, we have proposed a  model of ultra-cold bosons trapped in an optical lattice, that through periodic modulations of the lattice depth and interactions is capable of simulating at sufficiently low energy the physics of  dynamical gauge fields. We have investigated the low-energy excitations of a specific realization of our proposal  in the  resulting time-averaged dressed model. We have shown that these collective excitations are analogs, in specific regimes  to  free electromagnetic fields generated by gapless photons. We have sketched an experimental procedure to measure the resulting correlation functions corresponding to non-local operators in the original bosonic framework.
We believe that our present study will open a new route towards simulating gauge fields with bosonic and fermionic ultra-cold matter. We are currently working on  extensions to non-Abelian groups, higher and lower dimensions. We are also working on confirming via numerical Monte Carlo simulations and the recently developed tensor network simulations \cite{luca2,luca3, frank,mc,zoller} the details of the  phase diagram  of the Hamiltonian \eqref{eq:gaugeham} sketched in Fig.~\ref{figure2}.

\section{Acknowledgment} The authors would like to express a special thanks to the Mainz Institute for Theoretical Physics (MITP) for its hospitality and support.
O.D. and M.L. acknowledge support from EU grants OSYRIS (ERC-2013-AdG Grant No. 339106), SIQS (FP7-ICT-2011-9 No.600645), 
EQuaM (FP7/2007-2013 GrantNo. 323714), MINECO (FOQUS: FIS2013-46768-P, FISICATEAMO: FIS2016-79508-P, and Severo Ochoa Excellence Programme: SEV-2015-0522), Generalitat de
Catalunya grant AGAUR (2014 SGR 874 and CERCA/Programme), and Fundaci\'o Cellex. O.D., M.L. and J.Z. acknowledge support of QUIC (H2020-FETPROACT-2014 No.641122). O.D. and J.Z acknowledge support from National Science Center (Poland) projects No. DEC-2015/19/B/ST2/01028, and DEC-2016/21/B/ST2/01086 respectively.

\begin{appendix}
\section{Partition function of the $Z_N$ gauge system-weak coupling limit}
 \label{A_p}
In  the weak coupling limit, first we carry out the following transformation: $\hat{\mathcal{B}}_{p}\rightarrow \frac{2\pi}{N} \hat{\mathcal{B}}_{p}$ and $\frac{2\pi}{N} \hat{\mathcal{E}}_{(\mathbf j,\hat{x})} \rightarrow \hat{{\mathcal{E}}}_{(\mathbf j,\hat{x})}$ in the gauge Hamiltonian of Eq.\eqref{eq:gaugeham}, where $\hat{\mathcal{B}}_{p}$ has integer spectrum and the electric field operators have continuous compact spectrum with period $2\pi$. By using the Gauss Law: $\mathcal{E}_{(\mathbf j,\hat{x})}-\mathcal{E}_{(\mathbf j+\hat{x},\hat{x})}=-\mathcal{E}_{(\mathbf j,\hat{y})}+\mathcal{E}_{(\mathbf j+\hat{y},\hat{y})}$, we note that, $H_{\rm gauge}$ in Eq.\eqref{eq:gaugeham} becomes independent of $\mathcal{E}_{(\mathbf j,\hat{y})}$ operators. This permits us to write the magnetic field operator as: $\hat{\mathcal{B}}_{p}=\hat{\mathcal{A}}_{(\mathbf j+\hat{y},\hat{x})}-\hat{\mathcal{A}}_{(\mathbf j,\hat{x})}$, where on the same link : $\left [\hat{\mathcal{A}}, e^{i \hat{\mathcal{E}} } \right ] = e^{i \hat{\mathcal{E}}}$.  Taking the $U(1)$ limit with $N \rightarrow \infty $ and $g \rightarrow 0$ with fixed $Ng/2\pi$, $H_{\rm gauge}$ takes a form,
\begin{eqnarray}\label{weak}
	\frac{N^2}{4\pi^2 K } H_{\rm gauge } &=& \sum_{\mathbf j}  \left[\hat{\mathcal{A}}_{(\mathbf j+\hat{y},\hat{x})}-\hat{\mathcal{A}}_{(\mathbf j,\hat{x})}\right]^2 - 2\left[\frac{Ng}{2\pi}\right]^2 H_{\rm kin} \nonumber\\
	H_{\rm kin} &=& \sum_{\mathbf j} \left[ \cos\left ( 2\mathcal{E}_{(\mathbf j,\hat{x})}\right ) + 2\cos\left[\mathcal{E}_{(\mathbf j,\hat{x})}-\mathcal{E}_{(\mathbf j+\hat{y},\hat{x})}\right] \right. \nonumber \\
	&& \left.+   \cos\left[\mathcal{E}_{(\mathbf j,\hat{x})}-\mathcal{E}_{(\mathbf j+\hat{x},\hat{x})}\right] \right].
\end{eqnarray}
For simplicity, we denote $(\mathbf j,\hat{x}) \equiv j$ in this part. We will now find the partition function for Hamiltonian Eq.\eqref{weak} in Euclidean space-time and subsequently we introduce the imaginary time index $\tau$ with the new site index
$q \equiv (j,\tau)$. The partition function at finite temperature $1/\beta$ is given by,
${Z}_{\rm weak}=e^{-\beta H}=\sum^{N_{\tau}=\infty}_{N_{\tau}=-\infty} \Pi^{N_{\tau}}_1 \left \langle \{\mathcal{A}_{q}\}\right|e^{-\Delta_{\tau} H_{\rm kin}} e^{-\Delta_{\tau} H_{\rm pot}} \left | \{\mathcal{A}_{q+\hat{\tau}}\} \right \rangle $,
 where $N_{\tau}$ is the number of imaginary time steps of width $\Delta_{\tau}$ such that $N_{\tau}\Delta_{\tau}=2\pi\beta$. 
 The diagonal potential energy is given by, $H_{\rm pot}\left | \{{\mathcal{A}}_{q}\} \right \rangle = \sum_j \left[{\mathcal{A}}_{q+\hat{y}}-{\mathcal{A}}_{q}\right]^2 \left | \{{\mathcal{A}}_{q}\} \right \rangle$. While evaluating the partition function, we take the limit of $N_\tau \rightarrow \infty, \Delta_\tau \rightarrow 0$ such that $N_{\tau}\Delta_{\tau}=2\pi\beta$ remains constant. Using the standard periodic Gaussian approximation \cite{fradkin2}, the full partition function reads ${Z}=\Pi_{\tau} e^{-2\Delta_\tau H_{\rm pot}} Z_{\tau,\tau+1}$ with
\begin{eqnarray}
Z_{\tau,\tau+1} =  \sum^{\infty}_{\mathclap{\substack{m_{xq}=-\infty \\ m_{yj}=-\infty \\ l_j=-\infty }}} e^{ - \frac{2\pi^2 m^2_{xq}}{\tilde{g}}-\frac{2\pi^2 m^2_{yq}}{\tilde{g}}-\frac{2\pi^2l^2_j}{\tilde{g}}} &\times&  \left \langle \{\mathcal{A}_{q}\}\right| \nonumber\\
e^{i \mathcal{E}_{q}  \left[m_{xq}-m_{xq-\hat{x}}+m_{yq}-m_{yq-\hat{y}}+2l_q)\right] }\left | \{\mathcal{A}_{q+\hat{\tau}}\} \right \rangle, 
&&
\end{eqnarray}
where $m_{xq}, m_{yq}, l_q $ are integer-valued variables on Euclidean lattice site $q$
and $\tilde{g}=\Delta_\tau (gN)^2$. In the above expression we have neglected, for simplicity, the slight anisotropy along the $y$ direction. Additionally we introduce the symbolic 
differential action on a function $f_q$ as, $\partial_{\epsilon} f_q= f_q - f_{q-\hat{\epsilon}}$ where $\epsilon=x,y,\tau$ and a vector $\tilde{m}_q=(m_{xq},m_{yq},\mathcal{A}_{q})$. The commutation relation $\left [\hat{\mathcal{A}}, e^{i \hat{\mathcal{E}} } \right ] = e^{ i \hat{\mathcal{E}}}$ and the orthogonality of $\ket{\mathcal{A}_{q}}$ states leads to the constraint: $\nabla \cdot \tilde{m}_q=2l_q$. The total partition function reads,
\begin{eqnarray}
{Z}& = & \sum_{\mathclap{\substack{\{\tilde{m}_q\} \\ \{l_j\} }}} e^{ \sum_i \left[- \frac{2\pi^2}{\tilde{g}}\left(m^2_{xq}+m^2_{yq}+l^2_q\right)-2\Delta_\tau(\delta_y\mathcal{A}_{q})^2 \right]} \delta_{\nabla \cdot \tilde{m}_q,2l_q}, \nonumber\\ && \label{weakpart}
\end{eqnarray}
where $\delta_{m,n}$ is a Kronecker delta function. Next we use the identity $\int^{\infty}_{-\infty} \exp[-ax^2+bx]dx= C \exp[b^2/4a]$ to convert the sum of Gaussians over integers to continuous functions and also introduce a variable $\phi_q$ for the integral representation of Kronecker delta function $\delta_{\nabla \cdot \tilde{m}_q,2l_q}=\int^1_0 \exp[2\pi i (\nabla \cdot \tilde{m}_q-2l_q) \phi_q] d\phi_q$. The resulting expression is given by
\begin{eqnarray} \label{weak1}
Z &=& \sum_{\tilde{m}_q}\int D\theta_{x}D\theta_{y}D\theta_{\tau}\int D\phi Z_xZ_yZ_\tau \nonumber\\
&\times& \exp\left[\sum_q \left(-2\pi^2l^2_q/\tilde{g}+4\pi il_q \phi_q\right)\right], \nonumber\\
Z_x = \Pi_q&\exp&\left[-\tilde{g}\theta^2_{xq}+2\pi i m_{xq}(\theta_{xq}-\partial_x\phi_q)\right], \nonumber\\
Z_y = \Pi_q &\exp& \left[-\tilde{g}\theta^2_{yq}+2\pi i m_{yq}(\theta_{yq}-\partial_y\phi_q)\right], \nonumber\\
Z_{\tau}= \Pi_q&\exp&\left[-\pi^2\theta^2_{\tau q}/\Delta_{\tau}+2\pi i \mathcal{A}_q(\partial_y\theta_{\tau q}-\partial_\tau\phi_q)\right]\nonumber\\
D\theta_\epsilon &=& \Pi_q\int^{\infty}_{-\infty}d\theta_{\epsilon q}, \epsilon =x,y,\tau \nonumber\\
D\phi &=& \Pi_q \int^1_0 d\phi_q.
\end{eqnarray}
The invariance of the partition function under a constant shift $\phi_q\rightarrow \phi_q + c$ leads to the charge neutrality condition $\sum l_q=0$. After carrying out the summation over integer fields $\tilde{m}_q$ using the identity $\sum_n\exp[-i2\pi n x]=\sum_m \delta(m-x)$, we get,
\begin{eqnarray} \label{weak_2}
Z_\tau&=&\sum_{\{n_{\tau q}\}}\Pi_q\delta\left[n_{\tau q}-(\partial_y\theta_{\tau q}-\partial_\tau\phi_q)\right]\exp\left[-\pi^2\theta^2_{\tau q}/\Delta_{\tau}\right],\nonumber\\
Z_x&=&\sum_{\{n_{xq}\}}\Pi_j\delta\left[n_{xq}-(\theta_{xq}-\partial_x\phi_q)\right]\exp[-\tilde{g}\theta^2_{xq}], \nonumber\\
Z_y&=&\sum_{\{n_{yq}\}}\Pi_q\delta\left[n_{yq}-(\theta_{yq}-\partial_y\phi_q)\right]\exp[-\tilde{g}\theta^2_{yq}].
\end{eqnarray}
In the next step, we use the Helmholtz decomposition of the discrete vector field: $(n_{xq},n_{yq},n_{\tau q})=\vec{\nabla} n_{1 q} + \vec{n}_{\perp q}$ where $\nabla \cdot \vec{n}_{\perp q}=0$ and $n_{1 q}$ is a scalar integer field. Such a decomposition allows to redefine $\phi_q \rightarrow \phi_q + n_{1q}$ and accordingly the limit of integration transforms to $\sum_{n_{1q}}\int^1_0 d\phi_q \rightarrow \int^{\infty}_{-\infty} d\phi_q$. Moreover, to simplify the argument of $Z_{\tau}$, we introduce two new continuous variables $\varphi_q,\psi_q \in (-\infty,\infty)$ (also $D\phi=\Pi_q\int^{\infty}_{-\infty}\int^{\infty}_{-\infty} d\varphi_qd\psi_q$) such that, $\theta_{\tau q}=\partial_\tau\varphi_q+\partial_y\psi_q, \phi_q=\partial_y\varphi_q+\partial_\tau\psi_q$. In terms of the transformed variables, $Z_{\tau}=\sum_{\hat{\tau}\cdot\vec{n}_{\perp q}} \delta (\hat{\tau}\cdot\vec{n}_{\perp q}-\partial^2_{\tau}\psi_q+\partial^2_{y}\psi_q) \exp\left[-\pi^2(\partial_{\tau}\varphi_q+ \partial_y\psi_q)^2 /\
Delta_{\tau}\right]$. Therefore, the variable $\psi_q$ obeys $\partial^2_{\tau}\psi_q-\partial^2_{y}\psi_q=\hat{\tau}\cdot\vec{n}_{\perp q}$ with the solution: $\psi_q=\hat{\tau}\cdot\vec{n}_{\perp q}(\tau^2-q^2_y)$. Then, it is readily seen that for $\hat{\tau}\cdot\vec{n}_{\perp q} \neq 0$, the function $Z_{y}$ in Eq.\eqref{weak_2} is exponentially suppressed. This result can also be understood from noting that for long-wavelength fluctuations the original functions, $\partial_y\theta_{\tau q},\partial_\tau\phi_{q} \leq 1$, and as a result only the term with $\hat{\tau}\cdot\vec{n}_{\perp q}=0$ contributes significantly. The transverse component then satisfies $\vec{n}_{\perp q}=\vec{\nabla} \times \vec{a}_q, \partial_y a_{xq}=\partial_x a_{yq}$, where $a_q$ is a discrete vector field. In the gauge $\vec{\nabla} \cdot \vec{a}_q=0$, by carrying out the integration over $\theta_{xq},\theta_{yq}, \psi_q$ in Eq.\eqref{weak1}, \eqref{weak_2} leads to a simplified partition function, ${Z} = Z_{\rm 1} Z_{\rm 
dipole}$ with ${Z}_{1} = \sum_{\vec{a}_q}\exp\left[-\tilde{g}\sum_q(\vec{\nabla} \times \vec{a}_q)^2/2 \right]$ and
\begin{eqnarray}\label{finalZ}
Z_{\rm dipole} &=& \sum_{\{l_q\}}\Pi_q{\it z}_q \int D\varphi  \exp\left[ - \sum_q \Big[(\partial^2_{xy}\varphi_q)^2+(\partial^2_{yy}\varphi_q)^2 \right. \nonumber\\
&+& \left. \frac{\pi^2(\partial_\tau \varphi_q)^2}{\Delta_\tau \tilde{g}} + i \frac{4\pi}{\tilde{g}^{1/2}}\varphi_q\partial_y l_q \Big]\right] 
\end{eqnarray}
where we define an effective fugacity $z_q=\exp\left[-2\pi^2l^2_q/\tilde{g}\right]$. As we are interested in correlations involving functions like $\exp[-i\mathcal{E}]$ which only couple to the $\varphi$ fields, we ignore the contribution of $Z_1$ to the partition function. Next we take the limit of $N_\tau \rightarrow \infty, \Delta_\tau \rightarrow 0$ and carry out the Gaussian integral over $\varphi$, as a result the
final form of partition function reads
\begin{eqnarray}\label{classicalZ}
Z_{\rm dipole} &=& \sum_{\{l_q\}}\Pi_q{\it z}_q \exp\left[ - \sum_{qs} (l_q-l_{q-\hat{y}})\mathcal{G}(|q-s|)(l_s-l_{s-\hat{y}}) \right] \nonumber\\
&=& \sum_{\{l_q\}}\Pi_j{\it z}_q \exp\left[ - \sum_{qs} l_q\mathcal{G}_{\rm dipole}(|q-s|)l_s \right]
\end{eqnarray}
where for $q\equiv (j,\tau)$,
\begin{widetext}
	\begin{eqnarray}
	\mathcal{G}(q) &=& \lim_{\Delta_\tau \rightarrow 0} \frac{2}{\pi\tilde{g}}
	\int_{-\pi}^{\pi}\int_{-\pi}^{\pi} \int_{-\pi}^{\pi} \frac{e^{-ik_\tau\tau-ik_xj_x-ik_yj_y}dk_\tau dk_xdk_y}{16\sin^2k_x/2\sin^2k_y/2+16\sin^4k_y/2+\frac{4\pi^2}{\Delta_\tau\tilde{g}}\sin^2k_\tau/2} \nonumber\\
	\mathcal{G}_{\rm dipole}(q)&=& 2\mathcal{G}(q)-\mathcal{G}(q-\hat{y})-\mathcal{G}(q+\hat{y}).
	\end{eqnarray}
\end{widetext}
	The partition function in \eqref{classicalZ} is equivalent to a classical lattice gas of dipoles (denoted by integers $l_q$) interacting via $\mathcal{G}_{\rm dipole}(q)$
	and $\mathcal{G}(q)$ denotes the equivalent of Coulomb interaction for charges. The nature of $\mathcal{G}(q)$ and the form of the partition function suggests an absence of screening as in  three dimensional Coulomb gas. This can be seen best by assuming a dilute gas of dipoles due to fugacity $z_q$ and taking $l_q=\pm 1, 0$. After a suitable rescaling, we write Eq.\eqref{finalZ} in a modified Sine-Gordon like form in the continuum limit ($j_x \rightarrow x, j_y \rightarrow y, \tau \rightarrow \tau$): $ Z_{\rm dipole} \approx \int D\varphi e^{-S_{\rm dipole}}$ where
	\begin{eqnarray}
	S_{\rm dipole} &=&  \int d^3q \Big[(\partial^2_{xy}\varphi_q)^2+(\partial^2_{yy}\varphi_q)^2 
	+ (\partial_\tau \varphi_q)^2 \nonumber\\
	& - & z_0 \cos\left(\frac{4\pi^{1/2}}{(gN)^{1/4}}\partial_y\varphi_q\right)\Big], 
	\end{eqnarray}
	where we have introduced the fugacity $z_0=\pi\exp\left[-2\pi^2/\tilde{g}\right]/(\Delta_\tau\tilde{g})^{1/2}$ and $d^3q=dxdyd\tau$.
	
\section{Correlation function for the dipolar phase}
\label{A_c}

Now, we are in a position to calculate the correlation functions by evaluating $\mathcal{G}(q)$ (we remember that, $q$ denotes Euclidean space-time). For $(gN)<(gN)_c$, the 
charge-charge correlation function is given by: $\mathcal{C}(q) = \langle e^{i\varphi_q-i\varphi_0}\rangle = e^{-2\mathcal{G}(0)+2\mathcal{G}(q)} $ with
\begin{widetext}  
\begin{eqnarray}\label{intdip}
\mathcal{G}(q) &=& \frac{16}{\sqrt{gN}\pi^2}
 \int^{\infty}_{0}dk_\tau\int^{k_0}_{0}dk_y\left[\int^{k_0}_{0}dk_x  \frac{\cos(k_x x)\cos(k_y y)\cos(k_\tau \tau)}{k^2_\tau+k^4_y+k^2_yk^2_x}\right ] \nonumber\\
&\approx & \frac{16}{\sqrt{gN}\pi^2} \delta_{x,0}\int^{k_0}_{0}dk_y\int^{\infty}_{0}dk_\tau\frac{\tan^{-1}\left[\frac{k_0 k_y}{\sqrt{k^4_y+k^2_\tau}}\right]}{k_y\sqrt{k^4_y+k^2_\tau}}\cos(k_y y)\cos(k_\tau \tau) \\
&=& \frac{16}{\sqrt{gN}\pi^2}\delta_{x,0} \int^{k_0}_{0}dk_y\int^{\infty}_{0}dk_\tau\frac{\tan^{-1}\left[\frac{k_0k_y}{\sqrt{k^4_y+k^2_\tau}}\right]}{k_y\sqrt{k^4_y+k^2_\tau}}\left(\left[1-\cos(k_y y)\right]\left[1-\cos(k_\tau \tau)\right] - \left[1-\cos(k_\tau \tau)\right] - \left[1-\cos(k_y y)\right] + 1 \right). \nonumber
\end{eqnarray}
First, we carry out the $k_x$ integral inside brackets in the first line - we notice that the integral has a significant contribution for $x \ll 1/k_0, \cos(k_x x) \sim 1$. On the other hand, for $x\gg 1/k_0$ the integrand becomes highly oscillatory and thus have small contribution compare to $x \sim 0$ region. As a result, we approximate the integral as Kronecker delta function $\delta_{x,0}$. In the last line, due to the presence of terms like $(1-\cos k_y y)$ and $(1-\cos k_\tau \tau)$, the subsequent integrals have significant contributions from the momentum region $k_y \gtrsim 1/y, k_\tau \gtrsim 1/\tau$ for $y \gg 1, \tau \gg 1$. As a result, we rewrite the above expression as
\begin{eqnarray}\label{intdip1}
(\mathcal{G}(q) - \mathcal{G}(0)) &=& \frac{16}{\sqrt{gN}\pi^2} \delta_{x,0}\left[  \int^{k_0}_{1/y}\int^{\infty}_{1/\tau} - \int^{k_0}_{0}\int^{\infty}_{1/\tau}
- \int^{k_0}_{1/y}\int^{\infty}_{0}\right] \frac{\tan^{-1}\left[\frac{k_0k_y}{\sqrt{k^4_y+k^2_\tau}}\right]}{k_y\sqrt{k^4_y+k^2_\tau}} dk_y dk_\tau \nonumber\\
&=& \frac{16}{\sqrt{gN}\pi^2} \delta_{x,0}\left[  \int^{k_0}_{1/y}\int^{\pi/2}_{\tan^{-1}(1/\tau k^2_y)} - \int^{k_0}_{0}\int^{\pi/2}_{\tan^{-1}(1/\tau k^2_y)}
- \int^{k_0}_{1/y}\int^{\pi/2}_{0}\right] \frac{\tan^{-1}\left[\frac{k_0\cos \chi}{k_y}\right]}{k_y\cos \chi} dk_y d\chi
\end{eqnarray}
where in the second line we made the substitution $k_\tau=k^2_y \tan \chi $. Now we continue  considering the integral in the regime 
$\tau \gg y \gg \sqrt{\tau} \gg 1$. To do that, we first recast the last line in Eq.\eqref{intdip1} as,
\end{widetext}
\begin{eqnarray}
(\mathcal{G}(q) &-& \mathcal{G}(0))  =  \frac{16}{\sqrt{gN}\pi^2}\delta_{x,0}[I_1 + I_2], \nonumber\\
I_1 &=& - \int^{1/y}_0\int^{\pi/2}_{\tan^{-1}(1/\tau k^2_y)}\frac{\tan^{-1}\left[\frac{k_0\cos \chi}{k_y}\right]}{k_y\cos \chi}dk_y d\chi, \nonumber\\
I_2 &=& - \int^{k_0}_{1/y}\int^{\pi/2}_{0} \frac{\tan^{-1}\left[\frac{k_0\cos \chi}{k_y}\right]}{k_y\cos \chi} dk_y d\chi.
\end{eqnarray}
We concentrate on the evaluation of integral $I_1$. In the limit of $t$-integration, the lower limit $ \tan^{-1}(1/\tau k^2_y) \approx \pi/2 - \tau k^2_y $ as $k_y \rightarrow 1/y$ due to $y^2/\tau \gg 1$. On the other hand, for $k_y \rightarrow 0$, the lower limit approaches the upper limit of $\pi/2$ making the integral very small. By transforming $\pi/2-\chi \rightarrow \chi$ (accordingly $\cos\chi\rightarrow\sin \chi \sim \chi $) we get, 
\begin{eqnarray} \label{corint1}
I_1 &\approx& - \int^{1/y}_{0}\int^{\tau k^2_y}_{0}  \frac{\tan^{-1}\left[\frac{ k_0 \chi}{k_y}\right]}{k_y \chi} dk_yd\chi, \nonumber\\
&\stackrel{\chi/(\tau k^2_y)\rightarrow \chi }{\approx}& -\int^{1/y}_{0}\int^{1}_{0}  \frac{\tan^{-1}\left[{ \tau k_0k_y \chi}\right]}{k_y \chi} dk_yd\chi. 
\end{eqnarray}
In the last line of Eq.\eqref{corint1}, we again notice that the major contribution comes from the region $\chi \gtrsim 1/k_yk_0\tau, k_y \gtrsim 1/\tau $ and where $\tan^{-1}\left[{ \tau k_0k_y \chi}\right] \approx \pi/2$. As a result, the integration is expressed as,
\begin{eqnarray}
I_1&=&-\frac{\pi}{2}\int^{1/y}_{1/\tau}\int^{1}_{1/k_yk_0\tau} \frac{1}{k_y \chi} dk_yd\chi \nonumber\\
&=& -\frac{\pi}{2}\left[\log^2(k_0\tau/y)-\log^2(k_0)\right]\nonumber
\end{eqnarray}
We also carry out the integration for $I_2$ resulting in $I_2=-\frac{\pi}{2} \log^2(1/y)$. By joining both these results and from Eqs.\eqref{intdip},\eqref{intdip1}, we get the correlation function in the regime $\tau \gg y \gg \sqrt{\tau} \gg 1$:
\begin{widetext}
$$
\mathcal{C}(q) \sim \delta_{x,0} \exp\left[-\frac{8}{\pi\sqrt{gN}}\left(\log^2(k_0\tau/y)+\log^2(1/y)-\log^2(k_0)\right)\right] 
\sim \delta_{x,0} \left[\frac{1}{k_0\tau}\right]^{\frac{4\log(k_0\tau)}{\pi\sqrt{gN}}}\left[\frac{k_0\tau}{y^2}\right]^{\frac{4\log(y^2/k_0\tau)}{\pi\sqrt{gN}}}
$$
\end{widetext}
Similarly, we find the different regime for the charge correlation function $C(i)=\delta_{x,0} C(y,\tau)$, where
\begin{eqnarray}
C(y,\tau) &\sim& \left[\frac{1}{y}\right]^{\eta_3\log y }, y \gg \tau \gg 1 \nonumber\\
&\sim& \exp\left[-\eta_4 \frac{y^2}{\tau}y\right], \sqrt{\tau} \gg y \gg 1
\end{eqnarray}
 

\section{Relevant energy scales}
\label{A_r}
{\bf Optical lattice parameters :} For concreteness, we consider $^{133}$Cs as our {\it a}-bosons and $^{39}$K as our {\it b}-bosons. We use the inter-species resonance around $350$G \cite{patel} where the {\it a}-bosons interact strongly ($a_s\sim 2000$a$_0$) whereas {\it b}-bosons are essentially noninteracting ($a_s\sim 4$a$_0$). Our unit of energy is set by the recoil of {\it a}-bosons: $E_R=\pi^2\hbar^2/2m_a (\lambda/2)^2$ corresponding to a lattice constant of $\lambda/2$. Such a recoil energy is equivalent to a temperature scale of $100$nK for $\lambda=512$nm. First we fix the lattice parameters of the auxiliary boson lattice in \eqref{bosdepth} : $S=0.65, V^a_x=10E_R$. This corresponds to $J_{1x} \approx .12E_R$ and $J_{2x}/J_{1x} \approx 0.1$ which corresponds to a deep dimerized limit with {\it a}-bosons delocalized in the bond $s\hat{x}$ with odd $s_x$ with a gap $2J_1$. For half-filled auxiliary hard-core bosons, switching on the tunneling along y-direction will create two-leg ladders for each 
odd $s_x$. In such a ladder, hard-core bosons can be mapped to a spin-$1/2$ system and it may be shown that the dimer gap persists even for $J_y \sim J_1$ for half-filling. As a result, we can safely choose $J_y=0.2J_1$ corresponding to the lattice depth parameter $V^a_y \approx 7 E_R$. This sets the plaquette strength $K=.04 E_R$ which in turn sets the energy scale to observe the degenerate states.     

{\bf Parameters for A scheme:} This scheme is suitable for engineering the gauge Hamiltonian in U(1) limit, i.e $N \gg 1 (\alpha \rightarrow 0)$. The first constraint comes from the requirement that amplitude of tunneling driving will follow \eqref{constun} fixing the shaking amplitude in \eqref{shakamp_a} as $V_{sh} \approx 4.5 E_R$. Next, we choose our shaking frequency $\omega=5J_{1x}=0.6 E_R$. For {\it b} boson filling factor of $\bar{n}=20$ a lattice depth of $V^b\approx 7E_R$ is sufficient to reach coupling strength $g\sim 1$. Assuming Poissonian distribution for the number of particles at each site, we get an upper bound on the interaction shaking parameter $U_{\rm sh}/\omega \lesssim \alpha/\sqrt{\bar{n}}$. As a result, the phase strength $\alpha$ (for the chosen parameters and a z-direction trap frequency of $\Omega_{b} \approx 60$Hz) is given by $\alpha \approx 10^{-2}\delta a_s$, where $\delta a_s$ is the amplitude of shaking scattering length in nano-meters. $\alpha$ is limited by the constraint 
that $\alpha\sqrt{\bar{n}}\lesssim 1$ due to the validity of \eqref{approxphase}. 

{\bf Parameters for B scheme:} For scheme B, we are in the regime of lattice shaking with strong amplitude, $K/\omega \gg 1$. That constraints the duration of the experiment due to losses occurring by coupling to higher bands \cite{lacki13,andre}. Due to the separability of our lattice along the x- and y-directions, we can define a simpler one-dimensional model similar to the one in Eq.\eqref{shak_b} including the first excited band, 
to give a qualitative estimate of the loss rate. We consider   $H=H_{\rm tun} + H_{\rm on} + (H_{\rm t}+H_{\rm sp})\cos\omega t$, where
\begin{eqnarray}
H_{\rm tun} &=& -J^s_a\sum_{j}\left(\hat{s}^{\dagger}_{j} \hat{s}_{j+1} + h.c. \right)+J^p_a\sum_{j}\left(\hat{p}^{\dagger}_{j} \hat{p}_{j+1} + h.c. \right) \nonumber\\
&-& J_b\sum_{j} \left(\hat{b}^{\dagger}_{j} \hat{b}_{j+1} + h.c. \right)\nonumber \\
H_{\rm on} &=& U^{ss}_{ab} \sum_{j} \hat{n}^p_{j}\hat{n}_{bj}+U^{ps}_{ab} \sum_{j} \hat{n}^p_{aj}\hat{n}_{b{j}}+E_1\sum_{{j}} \hat{n}_{p{j}},\\
H_{\rm t} &=&  K^b_{\rm sh}\sum_{j} j \hat{n}_{b{j}}+K^a_{\rm sh}\sum_{j} j\left(\hat{n}^s_{aj} + \hat{n}^p_{aj}\right) \nonumber\\
H_{\rm sp} &=& K^a_{\rm sh 1}\sum_{j}\left(\hat{s}^{\dagger}_{j} \hat{p}_{j} + h.c. \right) \nonumber\\
&+& K^a_{\rm sh 2}\sum_{j}\left(\hat{p}^{\dagger}_{j} \hat{s}_{j+1} + \hat{s}^{\dagger}_{j} \hat{p}_{j+1} + h.c. \right), 
\end{eqnarray}
where $\hat{s}_j,\hat{s}^{\dagger}_{j};\hat{p}_{j},\hat{p}^{\dagger}_{j}$ are the annihilation and creation operators for the {\it a}-bosons in $s$-orbital and $p$-orbital at site $j$. Occupation number of the {\it a}-bosons is defined as $\hat{n}^s_{j}=\hat{s}^{\dagger}_{j}\hat{s}_{j};\hat{n}^p_{j}=\hat{p}^{\dagger}_{j}\hat{p}_{j}$. Interaction between lowest orbital {\it b}- and {\it a}-bosons in $\sigma$ orbitals is given by $U^{\sigma s}_{ab}$. $E_1$ is the energy of the $p$-orbital for the auxiliary bosons. Next, due to lattice shaking, on a single particle level, the $s$- and $p$-orbitals of the auxiliary bosons are coupled with coupling constants $K^a_{\rm sh}>K^a_{\rm sh 1}>K^a_{\rm sh2}$ due to the property of corresponding Wannier orbitals. We are in the regime of strong {\it a}-boson lattice shaking, i.e. $K^a_{\rm sh}/\omega \gg 1$ and weak {\it b}-boson shaking, $K^b_{\rm sh}/\omega \sim 1$. Moreover, we have assumed that the {\it b}-bosons do not get excited to the higher orbitals due to weak 
shaking with respect to the excitation energy.

Considering the resonant effect of interaction, we first apply the unitary transformation:  $\hat{U}=\exp [- { i} H_{\rm on}t -i H_{\rm t} \int^t_0 \cos\omega t' dt']$ which transfers time-dependence of the total Hamiltonian $H(t)$ into the tunneling amplitudes yielding $\tilde{H}=\hat{U}^\dagger H \hat{U} - {i}\hat{U}^\dagger [d_t \hat{U}]=\tilde{H}_{\rm tun}+\tilde{H}_{\rm sp}\cos\omega t$,
\begin{widetext}
\begin{eqnarray}\label{tavsh_b}
\tilde{H}_{\rm tun} &=& -J^s_a\sum_{j}\left(\hat{s}^{\dagger}_{j} \exp\left[-i U^{ss}_{ab}\hat{n}_{b{jx}} - i \frac{K^a_{\rm sh}}{\omega}\sin\omega t\right]\hat{s}_{j+1} + h.c. \right)+J^p_a\sum_{j}\left(\hat{p}^{\dagger}_{j}\exp\left[-i U^{ps}_{ab}\hat{n}_{b{jx}} - i \frac{K^a_{\rm sh}}{\omega}\sin\omega t\right] \hat{p}_{j+1} + h.c. \right) \nonumber\\
&-& J_b\sum_{j} \left(\hat{b}^{\dagger}_{j} \exp\left[-i U^{ss}_{ab}\hat{n}_{s{jx}} -i U^{ps}_{ab}\hat{n}_{p{jx}} - i \frac{K^b_{\rm sh}}{\omega}\sin\omega t\right]\hat{b}_{j+1} + h.c. \right)\nonumber \\
\tilde{H}_{\rm sp} &=& K^a_{\rm sh 1}\sum_{j}\left [ \hat{s}^{\dagger}_{j} \exp\left[-iE_1 t + i (U^{ss}_{ab}-U^{ps}_{ab})\hat{n}_{bj} t \right] \hat{p}_{j} +h.c.\right]\nonumber\\
&+& K^a_{\rm sh 2}\sum_{j}\left [ \hat{p}^{\dagger}_{j} \exp[iE_1 t + i (U^{ps}_{ab}\hat{n}_{bj}-U^{ss}_{ab}\hat{n}_{bj+1} )t - i \frac{K^a_{\rm sh}}{\omega}\sin\omega t]\hat{s}_{j+1}\right. \nonumber\\
&+& \left. \hat{s}^{\dagger}_{j} \exp[-iE_1 t + i (U^{ss}_{ab}\hat{n}_{bj}-U^{ps}_{ab}\hat{n}_{bj+1})t - i \frac{K^a_{\rm sh}}{\omega}\sin\omega t]\hat{p}_{j+1}  + h.c.\right], 
\end{eqnarray}
\end{widetext}
where the number difference is denoted by $n_{\sigma jx}=n_{\sigma j}-n_{\sigma j+1}$ where $\sigma$ is the boson species. The optical lattice parameter regime we are considering (same as scheme A), $U^{ss}_{ab}\approx 3U^{ps}_{ab}$, $U^{ss}_{ab}=2\omega$ and $K^a_{\rm sh 1}/K^a_{\rm sh}\approx 0.25$ and $ K^a_{\rm sh 2}/K^a_{\rm sh } \approx .05$. Moreover, the mean {\it b}-boson number at each site $\bar{n} \gg 1$. As a result we see that in the inter-orbital coupling Hamiltonian in Eq.\eqref{tavsh_b}, the first term is never resonant and its effect with in second-order perturbation theory leads to a loss rate:
$$
\Gamma_{\rm non} \propto \frac{(K^a_{\rm sh 1}/\omega)^2}{(E_1/\omega-4\bar{n}/3)^2}; K^a_{\rm sh 1}/\omega \ll |E_1/\omega-4\bar{n}/3|.     
$$
However, for the second term in $\tilde{H}_{\rm sp}$ in Eq.\eqref{tavsh_b}, an approximate resonant condition can be fulfilled with $E_1-4\bar{n}\omega/3 = N_{\rm res}\omega + \delta$ where $N_{\rm res} \gg 1$ is an integer and $\delta \ll \omega$ is the detuning. As a result, for resonant condition the loss is suppressed for time 
$$
t^{-1}_{\rm exp} \gtrsim {K^a_{\rm sh 2}}\mathcal{J}_{N_{\rm res}}(K^a_{\rm sh}/\omega).
$$
The above results show that, one can achieve a stable strong shaking regime by controlling the mean {\it b}-boson density.
 
Now we are in a position to give qualitative estimates for the loss rate and energy scales for scheme B. We choose  to study $\mathcal{Z}_3$ gauge theory for $\alpha=2\pi/3$. The most important constraint comes from maximum allowed shaking strength. For concreteness, we consider again the mean {\it b}-boson number $\bar{n} \sim 40$. Again assuming a Poissonian distribution, the boson number difference between neighboring sites ${n}_{b{i\delta}} \sim \pm 6$. We find that the asymptotic expansion of the Bessel function in Eq.\eqref{aympbess} remains valid for shaking strength $K^a_{\rm sh} \sim 25 \omega $. This shaking strength corresponds to a modified tunneling amplitude, $J_{1x} \approx .012 E_R$ and plaquette strength $K=.004 E_R$ which sets the energy gap for the deconfined phase. For similar lattice strengths as scheme-A, we get that $\Gamma_{\rm non} \sim .03$ and $1/t_{\rm exp} \gtrsim 10^{-5} E_R$, where the recoil energy $E_R$ is expressed in Hertz. The typical timescale to achieve the present 
experiment is also set by $K \sim \bar{n}J_b$, i.e $1/t_{\rm exp} \sim K^{-1} \sim 3\cdot10^{2}E_R$ which satisfy strongly the constrain for experimental time.

{\bf Effect of gauge invariance breaking due to the contact interaction between bosons}

One source of the gauge invariance breaking process originates from the contact interaction between the {\it b}-bosons with $U_{\rm bb}\neq 0$.
For $\bar{n}\gg 1$, the onsite interaction between bosons can be recast
as sum of two different term, i) the mean energy shift: $\propto \bar{n}^2 $, ii) and the gauge-breaking fluctuations: $H^U_{\rm break} \propto \frac{U_{\rm bb}}{2} \sum_{j} {\delta n}^2_{{j}}$,
where $\delta n=n_j-\bar{n}$ is the number fluctuation. A comparison with the plaquette operator suggests that gauge breaking part acts as a small perturbation for $|U_{\rm bb}|\bar{n}/2 \lesssim K$ where particle number fluctuations are assumed to obey Poissonian distribution, i.e. $\delta n_j \sim \sqrt{\bar{n}}$. This gives a stringent constraint for the scattering length of the bosons on order of few Bohr radii. {To get an estimate of the upper-limit for {\it b}-boson interaction, we use a Gaussian approximation for the boson wave-function which gives, $U_{\rm bb} = \sqrt{2/\pi}[V^b/E_R]^{1/2}[m_a/m_b][a_{\rm bb}/\sigma_z] E_R$, where we have introduced the  {\it b}-boson scattering length $a_{\rm bb}$, the z-direction confinement length $\sigma_z$. In the A scheme by using the parameters from Appendix \ref{A_r}, we get that the scattering length $a_{\rm bb} \lesssim 6a_0$. }  
\end{appendix}

\end{document}